\documentclass{jfm}
\usepackage{graphicx}
\usepackage{epstopdf, epsfig}
\usepackage{color}

\shorttitle{Effects of geometric confinement for a droplet between two parallel planes}
\shortauthor{C. Lv}

\title{\bf Effects of geometric confinement on a droplet between two parallel planes}

\author
 {
 Cunjing Lv\aff{1}
  \corresp{\email{lu@nmf.tu-darmstadt.de}}
  }

\affiliation
{
\aff{1}
Institute for Nano- and Microfluidics, Center of Smart Interfaces, Technische Universit{\"a}t Darmstadt, Alarich-Weiss-Stra{\ss}e 10, 64287 Darmstadt, Germany
}

\begin{document}

\maketitle

\begin{abstract}
When a droplet (which size is characterized by $R$) is confined between two parallel planes, its morphology will change accordingly to either varying the volume of the droplet or the separation (characterized by $ h $) between the planes. We are aiming at investigating how such a geometric confinement affects the wetting behaviours of a droplet. Our focus lies on two distinguished regimes: (1) a pancake shape in a Hele-Shaw cell when the droplet is highly compressed (i.e. $ h/R \ll 1 $), in which particular attention is paid on nonwetting and wetting cases, respectively; and (2) a liquid Hertzian contact rendered by a slight confinement (i.e. $h/(2R) \approx 1$) in a nonwetting case. To realize this aim, we first develop strict analytical expressions of the shape of the droplet which are available for arbitrary contact angles between the liquid and the solid planes, but in which the elliptic integrals indicate that these solutions are implicit. By employing asymptotic methods, we are able to give expressions of relevant geometrical and physical parameters (the Laplace pressure, droplet volume, surface energy and external force) in terms of sole functions of $R$ and $h$ in an explicit manner. Comparisons suggest that over a large range of $h/R$, our asymptotic results quantitatively agree well with the numerical solutions of the analytical expressions. This systematic study of the parameter space allows a comprehensive understanding of the geometric confinement  on wetting, especially a wide existence of logarithmic behaviours in a liquid Hertzian contact, to be identified.
\end{abstract}

% --------------------------------------------------------------------------------------------------
\section{Introduction}
% --
The cohesion of the fluid enables liquid droplets to minimize their surface area, resulting in spheres in the ideal case. However, when a droplet meets a confinement (can be specific geometric confinements, or external forces such as its own gravity, electric or magnetic fields), its shape will change accordingly. Meanwhile, ``capillary force'' caused by the meniscus of the droplet exerts on the confinement. Liquids can bridge solids through capillary force, which plays very important roles in a wide variety of natural creatures and artificial systems. For instance, many insects (such as the Asian weaver ants) that locomote by legs possess adhesive pads, which are rapidly releasable and adhesive force can be controlled during walking and running \citep{Federle2002, Peisker2012}; palm beetle uses wet adhesion to defend itself against attacking predators by adhering to the palm leaf using an array of liquid bridges \citep{Macner2014}; cat laps by a subtle mechanism based on water adhesion to the dorsal side of the tongue \citep{Reis2010}. For technological applications, even a small amount of liquid can produce strong enough cohesive force for building sand castles at the beach \citep{Mitarai2006}, and this adhesion effect between liquid and solid must be take into account in studies of powders, soil, and granular materials \citep{Butt2009}. Moreover, the precision of dip-pen nanolithography techniques also depends on the wetting behaviours of the substrate and the ink, as well as the geometric shape of the atomic force microscope tip \citep{Eichelsdoerfer2014}. In small scale, water can be absorbed at interfaces. With the miniaturization of many technological devices, such as microelectromechanical systems and heads on storage discs, it has become of primary importance to study capillary forces where they cause irreversible damaging stiction \citep{Riedo2002}. In all of these examples, the geometric confinement is crucial, which shapes the liquid, and accounts for the resulted static forces and adhesion behaviours.

% --
Thirty years ago, \citet{Park1984} carried out a pioneering work, in which  the pressure jump across the moving interface between two fluids confined in a Hele-Shaw cell was derived by employing asymptotic methods. This work sparked widely scientific interests in a broad range of fields about menisci confined in media \citep{Bostwick2015}, such as viscous fingering situations \citep{Bensimon1986, Geoffroy2006}, the Kelvin-Helmholtz instability \citep{Gondret1997, Plouraboue2002}, and translation and coalescence of droplets in bounded flows \citep{Lai2009, Zhu2016}. \citet{Paterson1995} experimentally studied the shapes of menisci distorted by wetting defects in Hele-Shaw cells and found the force exerted on the contact line. Very recently, \citet{Reyssat2015} promoted this topic to the wetting state of a liquid meniscus connecting a planar boundary and the surface of a horizontal cylinder placed above. Moreover, he not only analysed the dynamic wetting behaviours of drops and bubbles confined in wedges \citep{Reyssat2014}, but also studied capillary instability of two competing fluids in wedges \citep{Keiser2016}. These results, hence have been further exploited and widely applied as models to investigate relevant wetting phenomena in microfluidics \citep{Dangla2011, Dangla2013, Amselem2015}.

% --
In contrast, if a nonwetting droplet is exerted by a weak external stimulus, e.g., it is deformed by compressing two parallel planes slightly, or under its own gravity \citep{Lafuma2003, Srinivasan2011, Molacek2012}, the underlying physics is close to a Hertzian contact in elasticity \citep{Johnson1986} rather than the droplets in the Hele-Shaw cell. This case is particularly important for wetting on superhydrophobic materials because of their spectacular property of water repellency is realized through a very small contact region. Small deformations between the droplets and solids were checked by \citet{Mahadevan1999} on an inclined plane by rolling based on a scaling analysis. \citet{Chevy2012} promoted linear models using a logarithmic correction to describe the dynamics of a weakly deformed droplet that oscillates or bounces on superhydrophobic substrates, and existing experimental results support their theory. Moreover, \citet{Snoeijer2013} studied a heavily loaded, lubricated sliding contact between two elastic bodies in the flow and addressed a self-similar contact. The scaling laws and singularities revealed in the above works highly enrich our understandings of liquid Hertzian contacts. However, the investigation of the wetting state of droplets in a liquid Hertzian contact has not received much attention in the literature and is far from being fully understood. 

% --

The shape of the liquid is governed by the Laplace-Young equation. The challenge lies in the fact that the Laplace-Young equation cannot be solved analytically except in a few special cases \citep{Orr1975}. Therefore, efforts have been made to solve it numerically \citep{Macner2014}. An alternative approach is to approximate the liquid bridge shape with a priori assumed profile. In this view, seminal works were carried out by \citet{Haines1925} and \citet{Fisher1926} on soil mechanics, in which approaches were based on a toroidal approximation which involved treating the meridional profile of the liquid-vapor interface as an arc of a circle to estimate the capillary force. This leads to a simple closed-form solution, however, the toroidal approximation leads to a liquid bridge surface of nonconstant mean curvature which is inconsistent with the Laplace-Young equation. Consequently, there is no physically correct solution. To our knowledge, a comprehensive understanding of the effects of geometric confinement on a droplet with arbitrary contact angles, as well as arbitrary ratios between the size of the droplet and the distance of the planes, has remained very limited. All these facts motivate us to perform our current study. The objectives of the current work are as follows: (1) to obtain exact solutions of the shape of droplets confined between two planes, as well as other relevant geometrical and physical parameters, such as the curvature, pressure, surface energy, surface area, volume, capillary force; (2) considering these quantities must be expressed in terms of elliptic integrals without explicit solutions, we devote to express them as sole functions of some basic variables using asymptotic methods, which are easy to use and facilitates analyzing the relevant wetting phenomena; and (3) we examine the accuracy and applicability region of the explicit expressions, particularly the logarithmic behaviours in the liquid Hertzian contact which have not been sufficiently explored in the past. 

% --
The paper is structured as follows: We first obtain analytical expressions of the shape of the droplets for arbitrary contact angles (i.e. $\theta \in [0^\circ ~ 180^\circ]$) using strict theoretical derivations. Next, building on these analytical results, two distinguished regimes are particularly investigated using asymptotic methods: (1) a highly compressed droplet with a pancake shape in a Hele-Shaw cell in nonwetting and wetting cases; and (2) a liquid Hertzian contact rendered by a slight confinement in a nonwetting case, and this case is twofold: a constant distance of the two planes and a constant volume of the droplet. We give asymptotic expressions of relevant geometrical (i.e., the Laplace pressure, volume of the droplet) and physical parameters (i.e., the surface energy, capillary force exerted on the planes) in terms of $h$ and $x_2 $ only, denoting $ h $ the distance between the two planes and $ x_2 $ the maximum radius of the meniscus. Our asymptotical results quantitatively agree well with the theoretical results in a large range of $ x_2/h $. We close with a discussion and summary.
% --------------------------------------------------------------------------------------------------
%\clearpage
\section{Analytical study of a droplet confined by two parallel planes}
We consider a liquid droplet bridging two parallel planes. Not limited to a complete wetting state discussed previously \citep{deGennes1985, Ajaev2006, Bonn2009}, we will deal with the profiles of the droplets with arbitrary contact angles (as shown in figure \ref{fig:fig1} and \ref{fig:fig2}, respectively). Because the variation of $\theta$ and $h$ can strongly influence the wetting behaviours (e.g., Laplace pressure, capillary force, contact region), such an investigation is significant for a comprehensive understanding in a wide parameter space. It should be noted that forty years ago, \citet{Carroll1976} derived analytical solutions of the profile of a droplet-on-fiber systems in a hydrophilic wetting case, we will start to handle our problems employed the same method as in \citet{Carroll1976} but with necessary modifications related to the specific morphologies of the droplet-solid system in this study.
% --------------------------------------------------------------------------------------------------
\subsection{Hydrophobic wetting states}
% ************
\begin{figure}
  \centerline{\includegraphics[width=5.5cm]{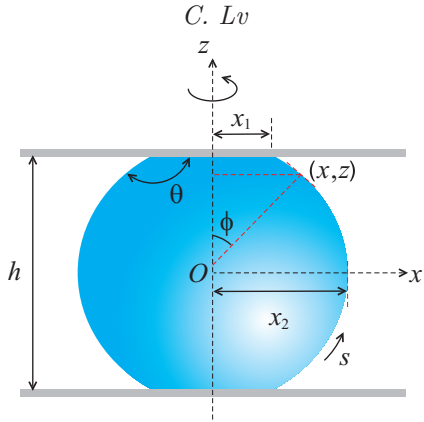}}
  \caption{Schematic defining geometrical parameters of a droplet confined by two parallel planes in 3-dimensional space without gravity, in which $ \theta \in [90^\circ ~ 180^\circ] $. An axisymmetric coordinate system is employed. $h$ is the distance between the two planes, $O$ is the center of the droplets and it is also the origin of the coordinate. $s$ represents the arc length of the profile, and $ \phi $ is the angle subtended by the normal at an arbitrary point $(x,z)$ of the meniscus to the axis of revolution (i.e. $z$-axis). $x_1$ is the solid-liquid contact radius and $x_2$ is the maximum value of the meniscus radius.}
\label{fig:fig1}
\end{figure}
% ************
We first consider a droplet in a hydrophobic wetting state, $\theta \in [90^\circ ~ 180^\circ]$, as shown in figure \ref{fig:fig1}. The equilibrium equation and the boundary condition are governed by the Laplace-Young equation $ \Delta P = 2H \gamma $ and the Young's equation $ \cos \theta = (\gamma_{\mathrm {SV}} - \gamma_{\mathrm {SL}})/\gamma $, respectively \citep{deGennes2004}, denoting $2H$, $\theta$, $ \gamma_{\mathrm {SV}} $, $ \gamma_{\mathrm {SL}} $, $ \gamma $ the total curvature of the meniscus, Young's contact angle, the solid-vapor, solid-liquid and liquid-vapor surface tensions. The Laplace pressure is defined by $ \Delta P = P_{\mathrm {in}} - P_{\mathrm {out}} $, in which $ P_{\mathrm {in}} $ and $ P_{\mathrm {out}} $ are the absolute pressures interior (liquid) and exterior (gas) to the droplet, respectively. In the absence of gravity, $ \Delta P $ and $2H$ are constant, and their values can be solely determined by $h$, $\theta$ and the droplet volume $V$. By employing an axisymmetric coordinate system and based on the differential geometry \citep{Struik1961}, the total curvature for an arbitrary point $ (x,z) $ at the meniscus can be expressed as, 
% ------------------
\begin{eqnarray}
2H = \frac{{\sin \phi }}{x} + \frac{{d\phi }}{{ds}},
\label{Eq_2_1}
\end{eqnarray}
% ------------------
% --
\noindent
in which $ \kappa_1 = \sin \phi / x $ and $ \kappa_2 = d \phi / ds $ are the azimuthal and longitudinal curvatures, and these two principal curvatures always have positive values in this case. $ x_{1} $ and $x_2$ are the radius of the solid-liquid contact region and the maximum radius of the meniscus. Eq. (\ref{Eq_2_1}) can be rewritten as $ 2Hx = {\mathrm d}(x \sin \phi) / {\mathrm d} x $, and a first integration with the boundary conditions (i.e., $ \phi_1 = \pi - \theta $ when $ x=x_1 $, and $ \phi_2 = \pi / 2 $ when $ x=x_2 $) gives,
% ------------------
\begin{eqnarray}
2H = \frac{{2\left( {{x_1}\sin \theta  - {x_2}} \right)}}{{x_1^2 - x_2^2}}.
\label{Eq_2_2}
\end{eqnarray}
% ------------------
% --
\noindent
On the basis of Carroll's method \citep{Carroll1976}, we obtain, 
% ------------------
\begin{eqnarray}
 - \frac{{dz}}{{dx}} = \frac{{{x^2} + a{x_1}{x_2}}}{{{{\left[ {\left( {x_2^2 - {x^2}} \right)\left( {{x^2} - {a^2}x_1^2} \right)} \right]}^{{1 \mathord{\left/
 {\vphantom {1 2}} \right.
 \kern-\nulldelimiterspace} 2}}}}},
\label{Eq_2_3}
\end{eqnarray}
% ------------------
% --
\noindent
in which $ a = (x_2 \sin \theta - x_1) / (x_2 - x_1 \sin \theta) $ is defined as a coefficient. By employing a transformation  $ x^{2} = x_{2}^{2}(1-{k^2}{{\sin}^2 \varphi} ) $ with $ k^2 = (x_2^2-{a^2}{x_1^2})/{x_2^2} $, Eq. (\ref{Eq_2_3}) can be integrated to give the exact solution of the droplet profile (i.e. $z=z(x)=z[x(\varphi)]$),
% ------------------
\begin{eqnarray}
z = \pm \left[ {a{x_1}F\left( {{\varphi},k} \right) + {x_2}E\left( {{\varphi},k} \right)} \right].
\label{Eq_2_4}
\end{eqnarray}
% ------------------
% --
\noindent
Here, $ F\left( {{\varphi},k} \right) $ and $ E\left( {{\varphi},k} \right) $ are incomplete elliptic integrals of the first and the second kinds, respectively. Furthermore, taking into consideration of the two boundary conditions: (1) $ {\phi _1} = \pi  - \theta  $, $ {\varphi _1} = \arcsin \sqrt {{{\left( {x_2^2 - x_1^2} \right)} \mathord{\left/
 {\vphantom {{\left( {x_2^2 - x_1^2} \right)} {{{\left( {k{x_2}} \right)}^2}}}} \right.
 \kern-\nulldelimiterspace} {{{\left( {k{x_2}} \right)}^2}}}} $ and $ z=h/2 $, when $ x=x_{1} $; (2) $ \phi_2 = \pi/2 $, $ {\varphi _2} = 0 $ and $ z=0 $, when $ x=x_{2} $, the height $h$, surface energy $U$ and volume $V$ of the droplet can be found,
% ------------------
\begin{eqnarray}
h = 2\left[ {a{x_1}F\left( {{\varphi _1},k} \right) + {x_2}E\left( {{\varphi _1},k} \right)} \right],
\label{Eq_2_5}
\end{eqnarray}
% ------------------
\begin{eqnarray}
U = \left[ 4\pi {x_2}\left( {a{x_1} + {x_2}} \right)E\left( {{\varphi _1},k} \right) - 2 \pi {x_1^2} \cos \theta \right] \gamma,
\label{Eq_2_6}
\end{eqnarray}
% ------------------
\begin{eqnarray}
\nonumber
 V & = \frac{{2\pi {x_2}}}{3} \left[ \left({2{a^2}x_1^2 + 3a{x_1}{x_2} + 2x_2^2}\right)E\left( {{\varphi _1},k} \right) - {a^2}x_1^2F\left( {{\varphi _1},k} \right) \right. \\ 
 &  + \frac{{x_1^2}}{{{x_2}}}\sqrt {\left( {x_2^2 - x_1^2} \right)\left( {1 - {a^2}} \right)} \left] \right.
\label{Eq_2_7}
\end{eqnarray}
% ------------------
\noindent
in which the surface energy is defined as $ U = (A_{\mathrm {LV}} - A_{\mathrm {SL}} \cos \theta) \gamma $ with the liquid-vapor and solid-liquid areas $ {A_{{\rm{LV}}}} = 4\pi {x_2}\left( {a{x_1} + {x_2}} \right)E\left( {{\varphi _1},k} \right) $ and $ A_{\rm SL} = 2 \pi {x_1^2} $, respectively.
% --------------------------------------------------------------------------------------------------
%\clearpage
\subsection{Hydrophilic wetting states}
% ************
\begin{figure}
  \centerline{\includegraphics[width=5.5cm]{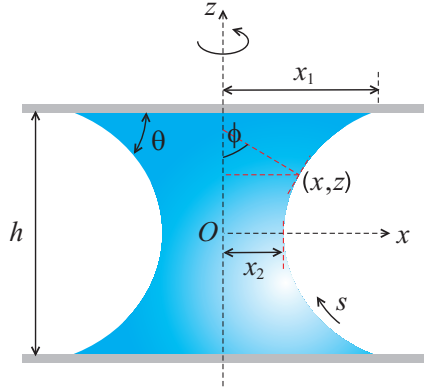}}
  \caption{A hydrophilic wetting state (i.e. $ \theta \in [0^\circ ~ 90^\circ] $) of a liquid bridge confined by two parallel planes, the definitions of the geometrical parameters are the same as shown in figure \ref{fig:fig1}.}
\label{fig:fig2}
\end{figure}
% ************
When the two planes are hydrophilic, $ \theta \in [0^\circ ~ 90^\circ] $, as shown in figure \ref{fig:fig2}, a liquid bridge will be formed, but in this case $ x_1 \geq x_2 $. The expression of the curvature $2H$ is the same as Eq. (\ref{Eq_2_1}). However, compared with Eq. (\ref{Eq_2_3}), the  equilibrium equation of the liquid meniscus has a very different formula,
% ------------------
\begin{eqnarray}
\frac{{dz}}{{dx}} = \frac{{a{x^2} + {x_1}{x_2}}}{{{{\left[ {\left( {{x^2} - x_2^2} \right)\left( {x_1^2 - {a^2}{x^2}} \right)} \right]}^{{1 \mathord{\left/
 {\vphantom {1 2}} \right.
 \kern-\nulldelimiterspace} 2}}}}}.
\label{Eq_2_8}
\end{eqnarray}
% ------------------
% --
\noindent
In this case $ a = (x_1\sin \theta  - x_2)/(x_1 - x_2\sin \theta) $. Another transformation $ (a x)^{2} = x_{1}^{2}(1-{k^2}{{\sin}^2 \varphi} ) $ with $ k^2 = (x_1^2-{a^2}{x_2^2})/{x_1^2} $ has to be employed. Then the two boundary conditions become: (1) $ {\phi _1} = \theta  $, $ {\varphi _1} = \arcsin \sqrt {{{\left( {1 - {a^2}} \right)} \mathord{\left/
 {\vphantom {{\left( {1 - {a^2}} \right)} {{k^2}}}} \right.
 \kern-\nulldelimiterspace} {{k^2}}}} $ and $ z = h/2 $, when $ x=x_{1} $; (2) $ \phi_2 = \pi/2 $, $ {\varphi _2} = \pi/2 $ and $ z = 0 $, when $ x=x_{2} $. Using these boundary conditions together with Eq. (\ref{Eq_2_8}), we can give the exact solution of the liquid meniscus (i.e. $z=z(x)=z[x(\varphi)]$),
% ------------------
\begin{eqnarray}
z = \pm \left. {\left[ {\frac{{{x_1}}}{a}E\left( {\varphi ,k} \right) + {x_2}F\left( {\varphi ,k} \right)} \right]} \right|_{{\varphi }}^{{\pi  \mathord{\left/
 {\vphantom {\pi  2}} \right.
 \kern-\nulldelimiterspace} 2}},
\label{Eq_2_9}
\end{eqnarray}
% ------------------
% --
\noindent
and the height $h$, surface energy $U$ and volume $V$ of the liquid bridge, 
% ------------------
\begin{eqnarray}
h = 2\left. {\left[ {\frac{{{x_1}}}{a}E\left( {\varphi ,k} \right) + {x_2}F\left( {\varphi ,k} \right)} \right]} \right|_{{\varphi _1}}^{{\pi  \mathord{\left/
 {\vphantom {\pi  2}} \right.
 \kern-\nulldelimiterspace} 2}},
\label{Eq_2_10}
\end{eqnarray}
% ------------------
\begin{eqnarray}
U = \left[ \left. {4\pi \frac{{{x_1}}}{a}\left( {\frac{{{x_1}}}{a} + {x_2}} \right)E\left( {\varphi ,k} \right)} \right|_{{\varphi _1}}^{{\pi  \mathord{\left/
 {\vphantom {\pi  2}} \right.
 \kern-\nulldelimiterspace} 2}} - 2 \pi {x_1^2} \cos \theta \right] \gamma,
\label{Eq_2_11}
\end{eqnarray}
% ------------------
\begin{eqnarray}
\nonumber
 V & = \frac{{2\pi }}{3}\left( {\frac{{{x_1}}}{a}} \right)\left\{ {\left[ {2{{\left( {\frac{{{x_1}}}{a}} \right)}^2} + 3\left( {\frac{{{x_1}}}{a}} \right){x_2} + 2x_2^2} \right]E\left( {\varphi ,k} \right) - x_2^2F\left( {\varphi ,k} \right)} \right. \\ 
 & \left. {\left. { + \left( {\frac{x}{{{x_1}}}} \right)\sqrt {x_1^2 - {{\left( {xa} \right)}^2}} \sqrt {{x^2} - x_2^2} } \right\}} \right|_{{\varphi _1}}^{{\pi  \mathord{\left/
 {\vphantom {\pi  2}} \right.
 \kern-\nulldelimiterspace} 2}} 
\label{Eq_2_12}
\end{eqnarray}
% ------------------
\noindent
in which $ A_{\rm{LV}} = 4 \pi (x_1/a) [(x_1/a) + x_2] E(\varphi,k) |_{\varphi_1}^{\pi / 2}$ and $ A_{\rm SL} = 2 \pi {x_1^2} $, respectively.
% --------------------------------------------------------------------------------------------------
%\clearpage
\subsection{Capillary forces}
We are interested in how much external force $F$ needed to apply (vertically) on the two planes to maintain the static wetting state. For the hydrophobic wetting state as shown in figure \ref{fig:fig1} and further schematically illustrated in figure \ref{fig:fig3}(e), on the one hand, we can carry out force analysis based on the balance of the solid-liquid contact area,
% ------------------
\begin{eqnarray}
P_{\mathrm {in}} \left( \pi x_1^2 \right) = P_{\mathrm {out}} \left( \pi x_1^2 \right) + 2 \pi x_1 \gamma \sin \theta + F,
\label{Eq_2_13}
\end{eqnarray}
% ------------------
\begin{eqnarray}
F = 2H \gamma \left( \pi x_1^2 \right) - 2 \pi x_1 \gamma \sin \theta
\label{Eq_2_14},
\end{eqnarray}
% ------------------
% --
\noindent
on the other hand, we can also carry out force analysis based on the balance of one half of the system (i.e., cutting horizontally in the middle of the liquid), 
% ------------------
\begin{eqnarray}
P_{\mathrm {in}} \left( \pi x_2^2 \right) = P_{\mathrm {out}} \left( \pi x_2^2 \right) + 2 \pi x_2 \gamma + F
\label{Eq_2_15}
\end{eqnarray}
% ------------------
\begin{eqnarray}
F = 2H \gamma \left( \pi x_2^2 \right) - 2 \pi x_2 \gamma,
\label{Eq_2_16}
\end{eqnarray}
% ------------------
% --
\noindent
It should be noted that when we combine Eq. (\ref{Eq_2_14}) and Eq. (\ref{Eq_2_16}) together, we can also get the value of $2H$, which is exactly the same we obtained (i.e. Eq. (\ref{Eq_2_2})) in the view of differential geometry \citep{Struik1961}. This result suggests an agreement of different parts in the above analyses. In fact, we can make force analysis based on the arbitrary cross section of the liquid, which will not change the conclusion. However, for the hydrophilic case (i.e. figure \ref{fig:fig2}), the sign of $F$ reverses, and pulling force instead of pushing force is required to maintain the static equilibrium of the system.
% --------------------------------------------------------------------------------------------------
%\clearpage
\section{Droplets in a Hele-Shaw cell}
In this section, we consider the situation when a droplet is highly confined between two parallel planes, i.e., $ h/x_2 \ll 1 $, so the droplet is in a Hele-Shaw cell. Considering the existence of the elliptic integrals, the analytical solutions we obtained in the last section are implicit, we are interested in giving explicit expressions of various geometrical and physical quantities using an asymptotic method. Next, we will particularly focus completely nonwetting and wetting cases to realize this aim.
% --------------------------------------------------------------------------------------------------
\subsection{Nonwetting states}
\begin{figure}
  \centerline{\includegraphics[width=13.0cm]{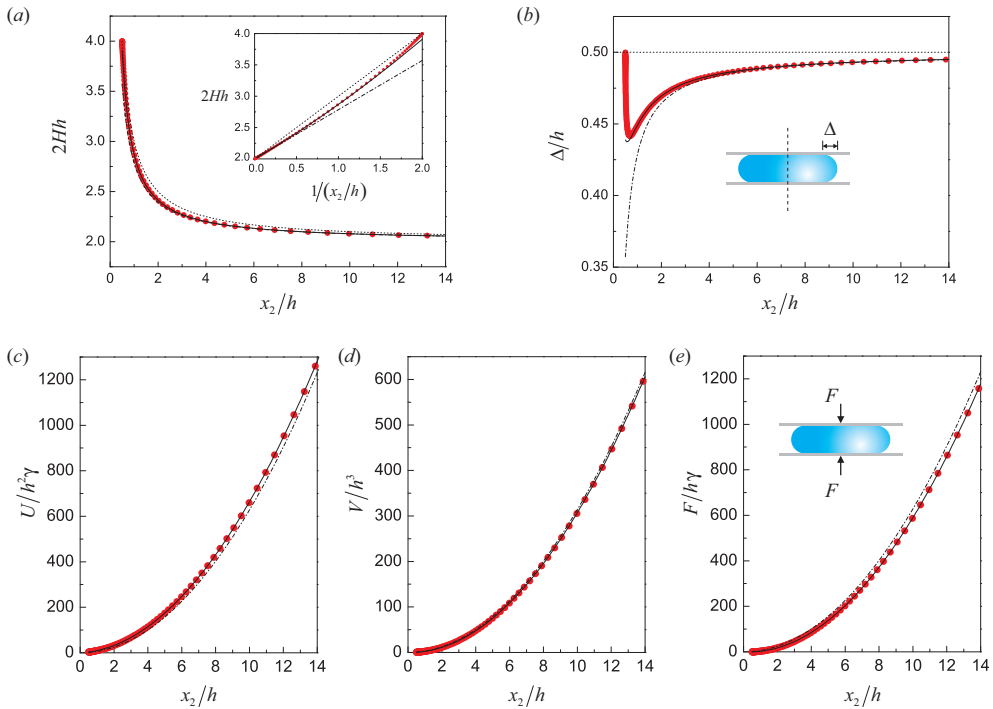}}
  \caption{Comparisons between the analytical and asymptotic results. The red dots are analytical results of Eq. (\ref{Eq_2_5}) - (\ref{Eq_2_7}) and Eq. (\ref{Eq_2_14}) solved numerically. ({\it a}) Relationship between the normalized values of $ 2 H $ and $ x_2 $. The inset gives the variation of $2H$ with one principle curvature $ 1/x_2 $ at the equator. The dotted, dashed and solid lines represent the normalized values of $ 2 H = 2/h + 1 / x_2 $, $ 2 H = 2/h + (\pi /4)(1/x_2) $ and Eq. (\ref{Eq_3_1}) up to the first order of $\varepsilon$, respectively. Here, $ x_2 /h \in [1/2, \infty] $ (or, $ 1/ (x_2/h) \in [0, 2] $). ({\it b}) Dependence of $ \Delta $ on $ x_2$. The red dots own a minimum point ($ x_2/h = 0.695 $, $ x_1/h = 0.253 $ and $ \Delta/h = 0.442 $), the dotted/dashed/solid line represents that we keep $\varepsilon$ up to the zero/first/second order in Eq. (\ref{Eq_3_4}). ({\it c})  $ U/h^2 \gamma $ {\it vs} $ x_{2}/h $; ({\it d}) $ V/h^3 $ {\it vs} $ x_2/h $. The dashed/solid line keeps $\varepsilon$ up to the zero/first order in Eq. (\ref{Eq_3_2}) and Eq. (\ref{Eq_3_3}), respectively. ({\it e}) $ F/h \gamma $ {\it vs} $ x_2/h $. The dashed and solid lines keep $\varepsilon$ up to the first ($\varepsilon^{-1}$) and second ($\varepsilon^{0}$) terms in Eq. (\ref{Eq_3_5}).}
\label{fig:fig3}
\end{figure}

We first consider a nonwetting state, i.e. $ \theta = 180^\circ $. Using power series, the handling of the elliptic integrals is straightforward (details are given in Appendix  \ref{appA_1}), and  we can derive the total curvature, surface energy and the volume of the droplet as functions of $h$ and $x_2$ only,
% ------------------
\begin{eqnarray}
2H = \frac{2}{h} + \frac{1}{{{x_2}}}\left[ {\frac{\pi }{4} + \left( \frac{\pi }{4} - \frac{\pi^2 }{16} \right)\varepsilon  +  \left( \frac{9 \pi}{32} - \frac{3 \pi^2}{16} + \frac{\pi^3}{32} \right) \varepsilon^2 + \cdots } \right],
\label{Eq_3_1}
\end{eqnarray}
% ------------------
\begin{eqnarray}
U = 2\pi x_2^2 \gamma \left[ { 1 + \left( {\pi  - 2} \right)\varepsilon  + \left( { 4 - \frac{\pi }{2} - \frac{{{\pi ^2}}}{4}} \right){\varepsilon ^2} +  \cdots } \right]  
\label{Eq_3_2}
\end{eqnarray}
% ------------------
\begin{eqnarray}
V = \pi x_2^2h\left[ {1 + \left( {\frac{\pi }{2} - 2} \right)\varepsilon  + \left( {\frac{8}{3} - \frac{{{\pi ^2}}}{4}} \right){\varepsilon ^2} +  \cdots } \right],
\label{Eq_3_3}
\end{eqnarray}
% ------------------
\noindent
in which, we define $ \varepsilon = (h/2)/x_{2} $ and $ \varepsilon \ll 1 $. Moreover, we use a parameter $\Delta$ to quantify the deformation of the droplet,
% ------------------
\begin{eqnarray}
\Delta = x_2 - x_1 = \frac{h}{2} \left[1 -  {\left( {\frac{\pi }{4} - \frac{1}{2}} \right)\varepsilon + \frac{ \left( \pi - 2 \right)^2 }{8} {\varepsilon ^2} +  \ldots } \right],
\label{Eq_3_4}
\end{eqnarray}
% ------------------
% --
\noindent
The expression of the Laplace pressure $\Delta P = 2H \gamma $ can also be obtained by Eq. (\ref{Eq_3_1}). The force (illustrated in the inset of figure \ref{fig:fig3}) applied on each plane is, 
% ------------------
\begin{eqnarray}
F = 2 \pi x_2 \gamma \left[ \frac{1}{2 \varepsilon} + \left( -1 + \frac{\pi}{8} \right) + \left( \frac{\pi}{8} - \frac{\pi^2}{32} \right) \varepsilon + \left( \frac{9}{64} \pi -\frac{3}{32} \pi^2 + \frac{1}{64} \pi^3 \right) \varepsilon^2 + \ldots \right].
\label{Eq_3_5}
\end{eqnarray}
% ------------------
% --
\noindent
Normalized by $ h $ and $ \gamma $, we define the dimensionless geometrical and physical parameters,
% ------------------
\begin{eqnarray}
2\bar H = 2H h,  \quad \bar U = U / h^2 \gamma, \quad \bar V = V / h^3, \quad \bar \Delta = \Delta / h, \quad \bar F = F / h \gamma. 
\label{Eq_3_6}
\end{eqnarray}
% ------------------
% --
which are for convenience given in appendix \ref{appA_1}. In order to figure out the scope of the availability using these power series, we solve Eqs. (\ref{Eq_2_5}) - (\ref{Eq_2_7}) numerically and make comparisons between them (red dots) and Eq. (\ref{Eq_3_1}) - Eq. (\ref{Eq_3_5}) up to different orders of $\varepsilon$ (black curves), as shown in figure \ref{fig:fig3}. Even though these results are obtained based on $h/x_2 \ll 1$, the comparisons suggest that they efficiently cover the theoretical solutions over a large range of $h/x_2$ (or, $x_2/h$).
% ************
% ************
% --------------------------------------------------------------------------------------------------
%\clearpage
\subsection{Wetting states}
\begin{figure}
  \centerline{\includegraphics[width=13.0cm]{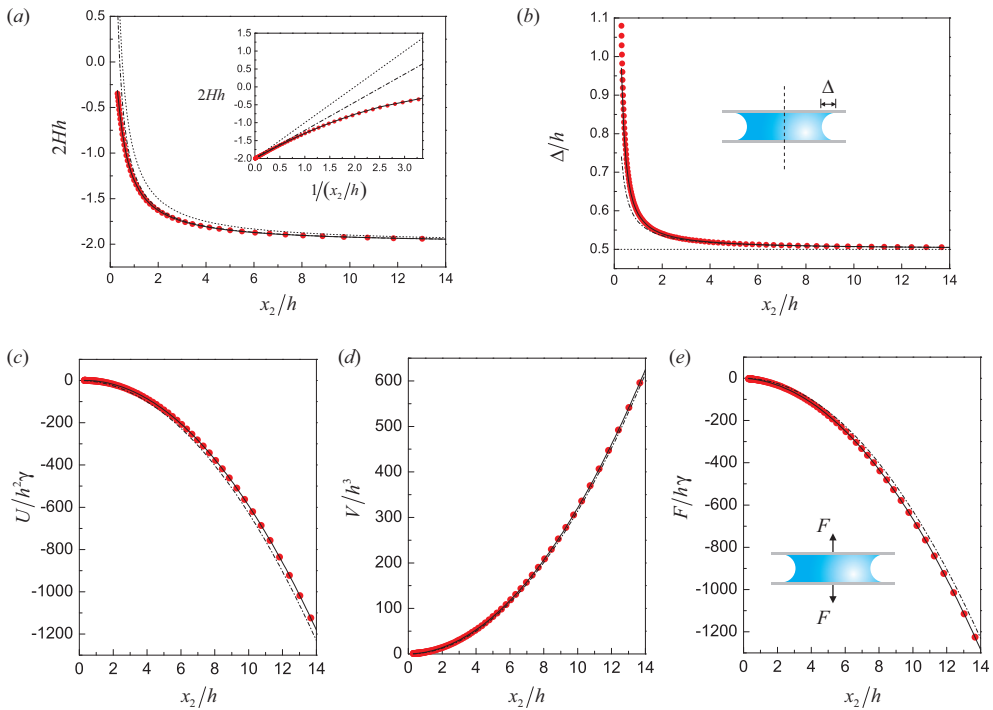}}
  \caption{Comparisons between the analytical and asymptotic results. The red dots are analytical results of Eq. (\ref{Eq_2_10}) - (\ref{Eq_2_12}) and Eq. (\ref{Eq_2_16}) solved numerically. ({\it a}) Relationship between the normalized values of $ 2H $ and $ x_2 $. The inset gives the variation of $2H$ with $ 1/x_2 $. The dotted, dashed and solid lines represent the normalized values of $ 2 H = -2 + 1 / x_2 $, $ 2 H = -2 + (\pi/4)(1/x_2)  $ and Eq. (\ref{Eq_3_7}) up to the first order of $\varepsilon$, respectively. Here, $ x_2/h \in [0.298, \infty] $ ( or $ 1/(x_2/h) \in [0, 3.361] $). ({\it b}) Dependence of $ \Delta $ on $x_2$. The dotted/dashed/solid line represents that we keep $\varepsilon$ up to the zero/first/second order in Eq. (\ref{Eq_3_10}). ({\it c})  $ U/h^2\gamma $ {\it vs} $ x_{2}/h $; ({\it d}) $ V/h^3 $ {\it vs} $ x_2/h $. The dashed/solid line keeps $\varepsilon$ up to the zero/first order in Eq. (\ref{Eq_3_8}) and Eq. (\ref{Eq_3_9}), respectively. ({\it e}) $ F/h \gamma $ {\it vs} $ x_2/h $. The dashed and solid lines keep $\varepsilon$ up to the first ($\varepsilon^{-1}$) and second ($\varepsilon^{0}$) terms in Eq. (\ref{Eq_3_11}). }
\label{fig:fig4}
\end{figure}

In this section, a nonwetting state, $ \theta = 0^\circ $, is considered. The total curvature, surface energy and volume of the droplet can also be expressed solely in terms of $h$ and $x_2$,
% ------------------
\begin{eqnarray}
2H = - \frac{2}{h} + \frac{1}{{{x_2}}}\left[ {\frac{\pi }{4} - \left( \frac{\pi }{4} - \frac{\pi^2 }{16} \right)\varepsilon  +  \left( \frac{9 \pi}{32} - \frac{3 \pi^2}{16} + \frac{\pi^3}{32} \right) \varepsilon^2 + \cdots } \right],
\label{Eq_3_7}
\end{eqnarray}
% ------------------
\begin{eqnarray}
U = 2\pi x_2^2 \gamma \left[ { - 1 + \left( {\pi  - 2} \right)\varepsilon  + \left( { - 4 + \frac{\pi }{2} + \frac{{{\pi ^2}}}{4}} \right){\varepsilon ^2} +  \cdots } \right] 
\label{Eq_3_8}
\end{eqnarray}
% ------------------
\begin{eqnarray}
V = \pi x_2^2h\left[ {1 + \left( { - \frac{\pi }{2} + 2} \right)\varepsilon  + \left( {\frac{8}{3} - \frac{{{\pi ^2}}}{4}} \right){\varepsilon ^2} +  \cdots } \right],
\label{Eq_3_9}
\end{eqnarray}
% ------------------
\begin{eqnarray}
\Delta = x_1 - x_2 = \frac{h}{2} \left[1 + {\left( {\frac{\pi }{4} - \frac{1}{2}} \right)\varepsilon  + \frac{\left( \pi - 2 \right)^2}{8} {\varepsilon ^2} +  \ldots } \right],
\label{Eq_3_10}
\end{eqnarray}
% ------------------
\begin{eqnarray}
F = 2 \pi x_2 \gamma \left[ - \frac{1}{2 \varepsilon} + \left( -1 + \frac{\pi}{8} \right) - \left( \frac{\pi}{8} - \frac{\pi^2}{32} \right) \varepsilon + \left( \frac{9}{64} \pi -\frac{3}{32} \pi^2 + \frac{1}{64} \pi^3 \right) \varepsilon^2 + \ldots \right].
\label{Eq_3_11}
\end{eqnarray}
% ------------------
\noindent
Using the same formula of Eq. (\ref{Eq_3_6}), we normalize Eq. (\ref{Eq_3_7}) - Eq. (\ref{Eq_3_11}) (see Appendix  \ref{appA_2}) and make comparisons between them (black curves) and the theoretical solutions (red dots) of (Eq. \ref{Eq_2_10}) - (\ref{Eq_2_12}), as shown in figure \ref{fig:fig4}.
% ************
% ************
% --------------------------------------------------------------------------------------------------
\subsection{Discussions}
We derive analytical solutions of the profiles of liquid droplets confined between two parallel planes. Because of the elliptic integrals (e.g. Eqs. (\ref{Eq_2_4}) - (\ref{Eq_2_7}) and Eqs. (\ref{Eq_2_9}) - (\ref{Eq_2_12})), we are not able to obtain explicit results directly. Giving the approximate solution using an asymptotic method to capture the essential physics is significant for a comprehensive understanding of the relevant wetting phenomena.

% --
In figure \ref{fig:fig3}(a) and \ref{fig:fig4}(a), we give comparisons to check the validity of the asymptotic results for the nonwetting and wetting cases, respectively. The solid lines represent that we consider $ \varepsilon $ to the first order in Eq. (\ref{Eq_3_1}) and  Eq. (\ref{Eq_3_7}). Form the inset of figure \ref{fig:fig3}(a), we can see this result is fairly accurate when $ 1/(x_2/h) \lesssim 1.5 $. It should be noted that figure \ref{fig:fig4}(a) suggests that for a complete wetting case, the precision of the asymptotic results with the first order of $\varepsilon$ is high globally. The dashed line means that we use $ 2H = \pm 2/h + (1/x_2) (\pi / 4) $ and the results are valid when $ 1/(x_2/h) \lesssim 0.5 $ and $ 1/(x_2/h) \lesssim 0.8 $ for the nonwetting and wetting states, respectively. Moreover, it is necessary to check the results using $ 2H = \pm 2/h + (1/x_2) $ (the dotted lines), because intuitively the two main curvatures are likely to be $ \pm 1/(h/2) $ and $ 1/x_2 $ when we specifically focus on the cross-section of the mid-height between the two planes. Unfortunately, this result is only acceptable when $1/(x_2/h) \ll 1$ due to the disappearance of one principal curvature ($1/x_2$), so we get $2H \approx 2/h $. The nonwetting case (i.e., the inset of figure \ref{fig:fig3}a) suggests that when $ x_2 / h \approx 2 $, this formula also approaches the analytical result, because in this case, the droplet is close to a sphere (i.e. $\kappa_1 \approx \kappa_2 \approx 2/h \approx 1/x_2 $). 

% --
It is interesting to note that in figure \ref{fig:fig3}(b), $\Delta = x_2 - x_1$ has a minimum ($ \Delta/h = 0.442 $, corresponding to $ x_2/h = 0.695 $ and $ x_1/h = 0.253 $). In fact, this critical point can be found theoretically, i.e., let $ \partial \Delta / \partial x_2 =0 $. Even though, due to the elliptic integrals, we have to employ numerical way to obtain the explicit values. Moreover, we have to emphasize that it is not efficient to use Eq. (\ref{Eq_3_4}) to approach $\Delta$ when $  x_2/h \to 0.5 $, because much higher orders of $\varepsilon$ have to be considered (we will address this question in the next section). For the case of completely wetting as shown in figure \ref{fig:fig4}(b), the scope of $\Delta = x_1 - x_2$ is $ \Delta / h \in [0.5, 1.080] $. $ \Delta_{\mathrm min} / h = 0.5 $ corresponds to $x_2/h \to \infty$ and $x_1/h \to \infty$; and $ \Delta_{\mathrm max} / h =1.080 $ corresponds to $ (x_2/h)_{\mathrm {min}} = 0.298 $ and $ (x_1/h)_{\mathrm {min}} = 1.377 $. Physically, the latter case means that the liquid bridge will break up into two parts under this critical condition. However, numerical methods have to be employed to solve the analytical expression to find these critical parameters. 

% --
Figure \ref{fig:fig3}(c)(d) and \ref{fig:fig4}(c)(d) indicates that the results expressed by power series are very acceptable compared to the analytical results (the red dots) when we take consideration of $\varepsilon$ up to the first order (solid lines) in Eq. (\ref{Eq_3_2}), Eq. (\ref{Eq_3_3}), Eq. (\ref{Eq_3_8}) and Eq. (\ref{Eq_3_9}) rather than just the zero order (the dashed lines). Furthermore, as shown in figure \ref{fig:fig3}(e) and \ref{fig:fig4}(e), the solid and dashed lines include $\varepsilon$ up to $\varepsilon^{-1}$ and $\varepsilon^{0}$ in Eq. (\ref{Eq_3_5}) and Eq. (\ref{Eq_3_11}), respectively. We could conclude that the predictions of $F = 2 \pi x_2 \gamma [\pm x_2/h + (-1+\pi/8)]$ are very acceptable compared with the red solids. These results we obtained have a clear physical picture. As one example and for the sake of simplicity, we carry out the following analysis: the surface energy can be approximately written as $ U \approx 2 \pi x_2^2 \gamma $ when $h/x_2 \ll 1$ for the nonwetting case, so the force applied on each plane is $ F \approx \Delta U / \Delta z \approx U/h \approx 2 \pi x_2^2 \gamma / h = \pi x_2 \gamma / \varepsilon$, which is exactly the same as the first term in Eq. (\ref{Eq_3_5}) (such analysis can also be applied to the wetting situation). The force diverges with a high ratio of $x_2/h$ (or small $\varepsilon$), this makes intuitive sense - the Laplace pressure is infinite.
% --------------------------------------------------------------------------------------------------
%\clearpage
\section{Liquid Hertzian contact}
In this section, we will discuss the situation when the droplet is slightly confined (i.e., $h/(2x_2) \to 1$) and in fact this situation is worth discussing from different views: (1) the distance $h$ between the two planes is fixed, which is a natural extension of what we discussed in the above. All the normalized parameters are still defined by Eq. (\ref{Eq_3_6}). This aspect is particularly meaningful for getting insights of water condensation in superhydrophobic microtextures with growing of the liquid \citep{Wang2015, Lv2015}, which has received substantial attention because of their ability to stay dry, self-clean and resist icing; (2) people would be also interested in how a droplet deforms with $h$ but with a certain volume, which suggests that a new characteristic length $r_0$ has to be chosen to normalize other geometrical parameters, denoting $r_0 = (3V/4 \pi)^{1/3}$ the radius of a spherical droplet with the same volume as the compressed droplet. The latter case is of significance for studying the superhydrophobicity of nonwetting materials \citep{Lafuma2003}, droplet impact on a rigid surface at small Weber numbers \citep{Chevy2012, Molacek2012}.
% ************
\begin{figure}
  \centerline{\includegraphics[width=5.5cm]{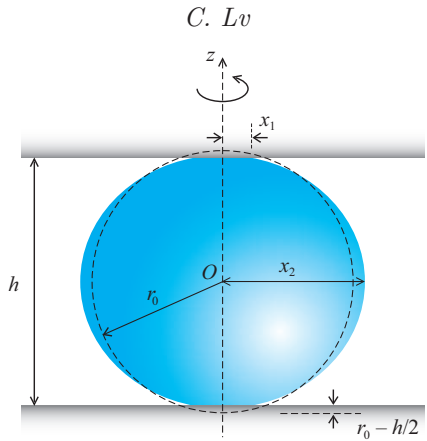}}
  \caption{Droplet confined slightly between two parallel planes in a liquid Hertzian contact, i.e., $ x_1/h \ll 1 $ and $ 2x_2/h \approx 2r_0/h \to 1 $. $ r_0 $ is the radius of a spherical droplet (indicated by the dashed line) with the same volume as the deformed droplet.}
\label{fig:fig5}
\end{figure}
% ************
% --------------------------------------------------------------------------------------------------
%\clearpage
\subsection{Constant distance $h$}
\begin{figure}
  \centerline{\includegraphics[width=13.0cm]{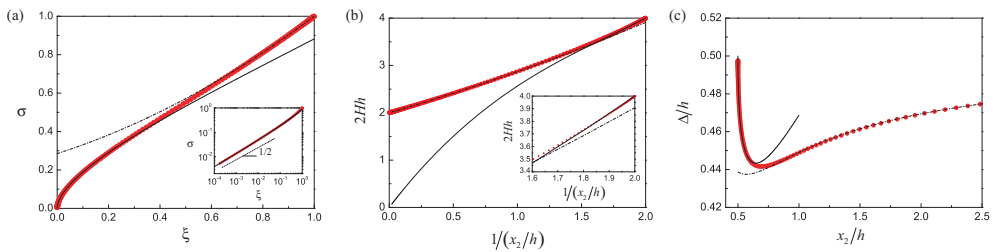}}
  \caption{Comparisons between analytical (red dots) and asymptotic results (solid curves). ({\it a}) The relationship between $\sigma$ and $\xi$ for the droplet in a liquid Hertzian contact (the solid line, Eq. (\ref{Eq_B6})) and for a pancake droplet (the dashed line, combing Eq. (\ref{Eq_4_1}) and (\ref{Eq_A5}), up to $\varepsilon^2$), respectively. The dashed line with a slope 1/2 in the inset is just used to guide the eye. ({\it b}) Relationship between the normalized curvature $ 2H $ and $ 1/x_2 $ in the full space (i.e. $ 1/(x_2/h) \in [0,2] $ and $ 2Hh \in [2,4] $) using Eq. (\ref{Eq_4_3}) (solid line) and Eq. (\ref{Eq_3_1}) (dashed line, up to $\varepsilon$), respectively. ({\it c}) The relationship between $ \Delta / h $ and $ x_2/h $ using Eq. (\ref{Eq_4_7}) (the solid line) and Eq. (\ref{Eq_3_4}) (the dashed line, up to $\varepsilon^2$), respectively.}
\label{fig:fig6}
\end{figure}

% --
In this case the droplet only has a very small contact region (i.e., $ x_1/h \ll 1 $), which makes the problem very different from a pancake shape in the Hele-Shaw cell as we discussed in the last section, and we are not able to tackle the elliptic integrals using the same method. An attempt to quantify the small deformation is let's first introduce another two infinitesimals with notations, 
% ------------------
\begin{eqnarray}
\sigma =\frac{x_{1}}{x_{2}}, ~~ \xi = 1 - \frac{h}{2x_2},
\label{Eq_4_1}
\end{eqnarray}
% ------------------
% --
\noindent
from which we know $\xi = 1 - \varepsilon$. Geometrically, these two parameters represent the relative deformations of the contact radius and the droplet size. We can quantify them as follows (see Appendix \ref{appB}),
% ------------------
\begin{eqnarray}
\xi = -\sigma ^{2} \ln \left( \frac{1}{2e^{1/2}} \sigma \right) + \frac{\sigma^4}{2} \ln \left( \frac{e^{1/4}}{2} \sigma \right) - \frac{\sigma^6}{4} \ln \left( \frac{e^{1/2}}{2} \sigma \right) \cdots
\label{Eq_4_2}
\end{eqnarray}
% ------------------
% --
\noindent
If we ignore the influence resulting from higher orders of $\sigma$, we can get $ \ln \xi \approx 2 \ln \sigma + \ln \ln (2e^{1/2}/\sigma) + \cdots \approx 2 \ln \sigma $. This suggests an approximate scaling relationship $ \xi \sim \sigma^\alpha $ with $\alpha \approx 2$ (in other words, $ x_1^2 \sim (x_2 - h/2) x_2 $), which is widely used in the contact problems between a droplet and a superhydrophobic surface \citep{Mahadevan1999, Aussillous2001}, and coalescence of droplets \citep{Eddi2013, Kavahpour2015} and air bubbles \citep{Thoroddsen2005}, based on some simple linear models. As shown in the inset of figure \ref{fig:fig6}, the analytical results (the red dots) are finely consistent with $ \xi \sim \sigma^2 $. However, to seek a higher precision, we have to take consideration of the logarithmic corrections in Eq. (\ref{Eq_4_2}), because in fact $\alpha$ can never exactly reach 2.  

% --
Since Eq. (\ref{Eq_4_2}) is a transcendental equation, we could not simply rewrite it using power series. By employing some asymptotic methods, we could give approximate solutions of $\sigma$ as the function of $\xi$ (see Appendix \ref{appB_1}), and thereby give explicit expressions of the following relationships as functions of $\xi$ (i.e., as functions solely of $x_2$ and $h$),
% ------------------
\begin{eqnarray}
2H = \frac{2}{x_2} \left\{ 1+ \xi \left[ -\ln \left( \frac{1}{2e^{1/2}} \xi^{1/2} \right) \right]^{ -1 + \frac{1}{2} \ln^{-1} \left( \frac{1}{2e^{1/2}} \xi^{1/2} \right) + \cdots } \right\} 
\label{Eq_4_3}
\end{eqnarray}
% ------------------
\begin{eqnarray}
U = 4\pi x_2^{2} \gamma \left\{ 1+ \xi \left[ -\ln \left( \frac{1}{2e^{1/2}} \xi^{1/2} \right) \right]^{ -1 + \frac{1}{2} \ln^{-1} \left( \frac{1}{2e^{1/2}} \xi^{1/2} \right) + \cdots } \right\} 
\label{Eq_4_4}
\end{eqnarray}
% ------------------
\begin{eqnarray}
V = \frac{4\pi }{3} x_2^3 \left\{ 1- \frac{3\xi}{2} \left[ -\ln \left( \frac{1}{2e^{1/2}} \xi^{1/2} \right) \right]^{ -1 + \frac{1}{2} \ln^{-1} \left( \frac{1}{2e^{1/2}} \xi^{1/2} \right) + \cdots } \right\} 
\label{Eq_4_5}
\end{eqnarray}
% ------------------
\begin{eqnarray}
F = 2 \pi x_2 \gamma \xi \left[ -\ln \left( \frac{1}{2e^{1/2}} \xi^{1/2} \right) \right]^{ -1 + \frac{1}{2} \ln^{-1} \left( \frac{1}{2e^{1/2}} \xi^{1/2} \right) + \cdots }
\label{Eq_4_6}
\end{eqnarray}
% ------------------
\begin{eqnarray}
\Delta = x_2 \left\{ 1 - \xi^{1/2} \cdot \left[ - \ln \left( \frac{1}{2e^{1/2}} \xi^{1/2} \right) \right]^{ - \frac{1}{2} + \frac{1}{4} \ln^{-1} (\frac{1}{2e^{1/2}} \xi^{1/2}) + \cdots } \right\}
\label{Eq_4_7}
\end{eqnarray}
% ------------------
% --
If we ignore the infinitesimals of $\xi$, we can obtain $ 2H \approx 2/x_2 $, $ U \approx 4 \pi x_2^2 \gamma $, $ V \approx 4 \pi x_2^3 /3 $, $ F \approx 0 $ and $ \Delta \approx x_2 $, which can also be obtained intuitively due to the small deformation of the droplet. Eq. (\ref{Eq_4_3}) - (\ref{Eq_4_7}) indicates that these relationships have logarithmic behaviours. In figure \ref{fig:figS1} (see Appendix \ref{appB_1}), we give more comparisons between asymptotic and analytical results.
% ************
% ************
% --------------------------------------------------------------------------------------------------
%\clearpage
\subsection{Constant volume $V$}
% --
For the droplet with a constant volume $V = 4 \pi r_0^3 /3$, we check how other relationships vary as functions of $x_2$. In this case, $r_0$ and $\gamma$ are chosen to normalize the geometrical and physical parameters,
% ------------------
\begin{eqnarray}
\hat x_2 = x_2 / r_0, \quad \hat h = h / r_0, \quad \hat \Delta = \Delta / r_0, \quad 2\hat H = 2H r_0,  \quad \hat U = U / r_0^2 \gamma, \quad  \hat F = F / r_0 \gamma. 
\label{Eq_4_8}
\end{eqnarray}
% ------------------
% --
\noindent
On the basis of $(\hat x_2 - 1) \to 0$, we can get (details are given in Appendix \ref{appB_2}),
% ------------------
\begin{eqnarray}
h = 2 x_2 \left[ 1 + \left( \hat x_2 - 1 \right) \ln \left( \frac{\hat x_2 - 1}{2e} \right) - \frac{3}{2} \left( \hat x_2 - 1 \right)^2 \ln \left( \frac{\hat x_2 - 1}{2e^{-1/3}} \right) + \cdots \right],
\label{Eq_4_9}
\end{eqnarray}
% ------------------
\begin{eqnarray}
2H = \frac{2}{x_2} \left[ 1+ 2 \left( \hat x_2 - 1 \right) + 3 \left( \hat x_2 - 1 \right)^2 + \cdots \right],
\label{Eq_4_10}
\end{eqnarray}
% ------------------
\begin{eqnarray}
U = 4\pi x_2^2 \gamma \left[ 1 - 2 \left( \hat x_2 - 1 \right) - \left( \hat x_2 - 1 \right)^2 \ln \left( \frac{\hat x_2 - 1}{2e^{5/2}} \right) + \cdots \right],
\label{Eq_4_11}
\end{eqnarray}
% ------------------
\begin{eqnarray}
F = 4 \pi x_2 \gamma \left[ \left( \hat x_2 - 1 \right) + \frac{3}{2} \left( \hat x_2 - 1 \right)^2 + \cdots \right],
\label{Eq_4_12}
\end{eqnarray}
% ------------------
\begin{eqnarray}
\Delta = x_2 \left[ 1 - \sqrt{2} \left( \hat x_2 - 1 \right)^{1/2} + \frac{\sqrt 2}{4} \left( \hat x_2 - 1 \right)^{3/2} + \cdots \right]. 
\label{Eq_4_13}
\end{eqnarray}
% ------------------
% --
\noindent
Details about the comparisons between these equations and analytical solutions are given in figure \ref{fig:figS2} (see Appendix  \ref{appB_2}). These equations can degrade to $ h \approx 2 r_0 $, $ 2H \approx 2/r_0 $, $ U \approx 4 \pi r_0^2 \gamma $, $ V \approx 4 \pi r_0^3 /3 $, $ F \approx 0 $ and $ \Delta \approx x_2 $ if we ignore the infinitesimals of $(\hat x_2 - 1)$.

% --
Similar to the last section, we are also interested at how the contact region deforms. Let's define $\Omega = r_0 - h/2$ and $ \hat \Omega = \Omega / r_0 $ as shown in figure \ref{fig:fig5}. As a consequence, we can get $\Omega x_2 \sim x_1^2$ and $ \Omega r_0 \sim x_1^2 $ as shown in figure \ref{fig:fig7}(a) in dimensionless form (see Appendix \ref{appB_2}). What's more, figure \ref{fig:fig7} also includes that when a droplet has a pancake shape (i.e. $\hat x_1 \gg 1$), $\hat \Omega \hat x_2 \sim \hat x_1$ and $\hat \Omega \sim \hat x_1^0$ because $\Omega = r_0 - h/2 \approx r_0$. We also give the relationship between $h$ and $x_2$ in the dimensionless form in the inset of figure \ref{fig:fig7}(b). Furthermore, figure \ref{fig:fig7} indicates a logarithmic relationship between $h$ and $x_2$ in the semi-log plot. And this scaling relationship can be directly obtained from Eq. (\ref{Eq_4_9}), which indicates the influence of higher orders of $\mathcal{O} \left( (\hat x_2 -1)^2 \right)$ has a weak effect on this logarithmic behaviour.
% ************
\begin{figure}
  \centerline{\includegraphics[width=13.0cm]{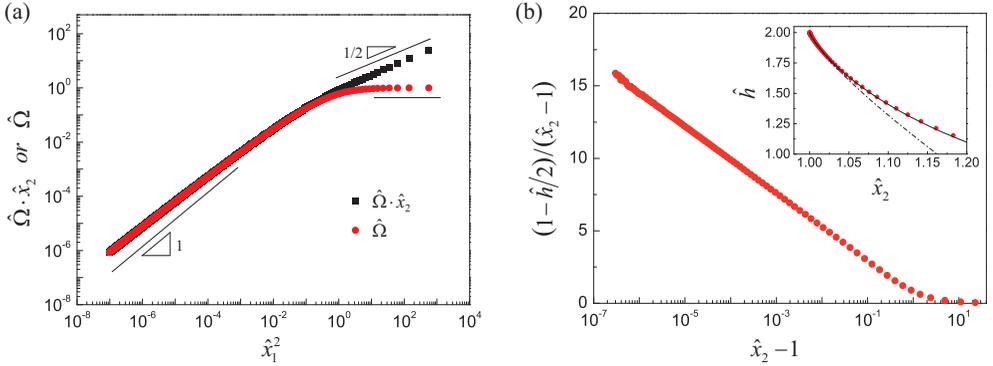}}
  \caption{({\it a}) The dependence of $ \Omega x_2 $ on $ x_1^2 $ (black squares) and $ \Omega r_0 $ on $ x_1^2 $ (red dots) in dimensionless form. ({\it b}) A logarithmic law $ (r_0 - h/2) \sim (x_2 - r_0) \ln (x_2 - r_0)$ is extracted from Eq. (\ref{Eq_4_9}) and shown in the semi-log plot in dimensionless form. The inset gives the relationship between $ h $ and $ x_2 $.}
\label{fig:fig7}
\end{figure}
% ************
% --------------------------------------------------------------------------------------------------
%\clearpage
\subsection{Discussions}
In this section, we systematically investigate the wetting behaviours of liquid Hertzian contacts in two cases. Because of the high nonlinear properties of the elliptic integrals and the transcendental equations produced, we cannot deal with them using power series as we used for a pancake droplet. Fortunately, on the basis of suitable asymptotic methods, we can obtain approximations explicitly, and they could reach the exact values with a high precision with proper terms. Very different from the linear models used previously \citep{Mahadevan1999}, our results show that singularities widely exist in these relationships and most of them obey logarithmic behaviours. 

% --
In the view of physics, since in the liquid Hertzian contact, the relationship between the exerted force and the deformation of the droplet is important, it is worth expressing $F$ as functions of the droplet deformation. If we ignore $\mathcal{O} (\sigma^4)$, in other words, let's $F \approx 2 \pi x_2 \gamma \sigma^2$ (see Eq. (\ref{Eq_B10}) and (\ref{Eq_B21})), we get, 
% ------------------
\begin{eqnarray}
\left. x_2 - \frac{h}{2} \right|_{h = const.} \approx - x_2 \sigma^2 \ln \frac{\sigma}{2e^{1/2}} \approx - \frac{F}{4 \pi \gamma} \ln \frac{F}{8 \pi e x_2 \gamma},
\label{Eq_4_14}
\end{eqnarray}
% ------------------
% ------------------
\begin{eqnarray}
\left. r_0 - \frac{h}{2} \right|_{V = const.} \approx -r_0 \sigma^2 \ln \frac{\sigma}{2} \approx - \frac{F}{4 \pi \gamma} \ln \frac{F}{8 \pi r_0 \gamma}.
\label{Eq_4_15}
\end{eqnarray}
% ------------------
% --
\noindent
in which we use $(x_2 - h/2)$ and $(r_0 - h/2)$ to characterize the deformation of the droplet in the vertical direction in these two different cases. We have to emphasize that the underlying mechanisms behind Eq. (\ref{Eq_4_14}) and Eq. (\ref{Eq_4_15}) are the same.

% --
In order to check the validity of the above results, we explore them further in a different view: if the droplet with a constant volume deformed under its own gravity, i.e. $F = Mg = 4 \pi r_0^3 \kappa^2 \gamma /3$, denoting $\kappa^{-1}=\sqrt{\gamma/ \rho g} $, $M$, $\rho$ and $g$ the capillary length \citep{deGennes2004}, mass, mass density and gravitational acceleration, Eq. (\ref{Eq_4_15}) leads to, 
% ------------------
\begin{eqnarray}
\left. r_0 - \frac{h}{2} \right|_{V = const.} \approx - \frac{r_0^3 \kappa^2}{3} \ln \left( \frac{r_0^2 \kappa^2}{6} \right).
\label{Eq_4_16}
\end{eqnarray}
% ------------------
% --
\noindent
Interestingly, Eq. (\ref{Eq_4_16}) is exactly the same result derived by \citet{Chevy2012} using a completely different method in the study of the liquid Hertzian contact, which was verified experimentally as well. Not limited to the specific problem we focused on, we envision our conclusions, e.g. Eq. (\ref{Eq_4_14}) and (\ref{Eq_4_15}), are more general and could be used to predict the deformation of a droplet in other types of force fields, for instance, magnetic force, electric force, droplets under acceleration, etc.
% --------------------------------------------------------------------------------------------------
\section{Concluding remarks}
Wetting behaviours of a droplet under the confinement of two parallel rigid planes have been systematically studied. Our results have shown the following.

% --
(1) Analytical expressions for a droplet confined between two parallel planes are obtained. These solutions are available for arbitrary contact angles. However, because of the elliptic integral, we cannot get their exact solutions, numerical methods have to be employed to solve them; (2) For droplets with a pancake shape in a Hele-Shaw cell, employing power series, we give the approximate expressions of the curvature, surface energy, volume, capillary force as functions of $x_2$ and $h$ for a nonwetting and wetting case; (3) For nonwetting droplets in a liquid Hertzian contact, two cases are discussed. Due to the intrinsically complicated transcendental equations, we cannot easily obtain the explicit expression as we were using power series. However, asymptotic methods can still be used to obtain approximate solutions of the relevant geometrical and physical parameters as functions of the very basic variables of the system, i.e. $x_2$ and $h$ (or $r_0$); (4) These explicit results we obtained are not only significantly simpler than the analytical expressions, but also reflect essential features of the wetting behaviours in corresponding regimes. Moreover, the procedures used to drive these results can be easily extended up to higher orders to satisfy the desired accuracy, and cover a larger regime of $h/x_2$ as well. 

% --
These findings suggest a particular avenue of enquiry deserving of further attention: (1) As an extension, our theory (Sec. 2.2) can be used to find the critical diameter/volume of a liquid bridge between two hydrophilic planes; (2) We envision that our methods could be extended to handle droplet problems with other value of contact angles, for droplets both in a Hele-Shaw cell or in a state of liquid Hertzian contact; (3) We hope our study are not limited to specific shapes of the substrates but motivate further exploration of interactions between droplets/bubbles and substrates with various configurations (e.g., two spheres, one plane and one sphere, substrates with conical/cylindrical ends and a plane) based on the basic ideas we put forward in this work. 

% --
From a practical point of view, for future studies, it is recommended to take consideration of other influences, such as contact line pinning and line tension \citep{deGennes1985, Amirfazli2004}, van der Waals' force \citep{Israelachvili2011}, long-range attractive forces \citep{Christenson1988}, elasticity of the substrate \citep{Roman2010}, etc. More importantly, carrying out laboratory experiments to check the theories presented in this paper is our next step.
% --------------------------------------------------------------------------------------------------
\section*{Acknowledgements}
The support of the Alexander von Humboldt Foundation is gratefully acknowledged. 
% --------------------------------------------------------------------------------------------------
\appendix
% --------------------------------------------------------------------------------------------------
\section{Deductions of the asymptotic solution of a droplet in a Hele-Shaw cell}\label{appA}
In the following, we will give explicit solutions of a droplet in a Hele-Shaw cell, in which the power series will be employed. 
% --------------------------------------------------------------------------------------------------
\subsection{Nonwetting states}\label{appA_1}
If the Young's contact angle is $ \theta = 180^{\circ} $, we get $ a = -x_{1}/x_{2} $ and $ k^{2} = 1-(x_{1}/x_{2})^{4} $. When $h / x_{2} \ll 1$, the key step is let $ \Delta = x_{2} - x_{1} $ and $ \delta = \Delta / x_{2} $. Herein, $ a=-1+\delta $ and $ k = \sqrt{1-(1-\delta)^4} $. We know $\Delta \to h/2$, so $ \delta \ll 1 $. Correspondingly, $\delta$ is employed as an infinitesimal to make Taylor series for the relevant quantities,
% ------------------
\begin{eqnarray}
\varphi_1 = \frac{\pi}{4} + \frac{1}{2} \delta + \frac{1}{4} \delta^2 + \frac{1}{12} \delta^3 + \ldots,
\label{Eq_A1}
\end{eqnarray}
% ------------------
\begin{eqnarray}
F\left( {{\varphi _1},k} \right) = \frac{\pi }{4} + \frac{\pi }{4}\delta  + \frac{{3\pi }}{{16}}{\delta ^2} + \frac{\pi }{8}{\delta ^3} + \ldots,
\label{Eq_A2}
\end{eqnarray}
% ------------------
\begin{eqnarray}
E\left( {{\varphi _1},k} \right) = \frac{\pi }{4} + \left( {1 - \frac{\pi }{4}} \right)\delta  + \left( { - \frac{1}{2} + \frac{{3\pi }}{{16}}} \right){\delta ^2} - \frac{{3\pi }}{{256}}{\delta ^4} +  \ldots 
\label{Eq_A3}
\end{eqnarray}
% ------------------
% --
\noindent
Substitute Eq. (\ref{Eq_A1}) - (\ref{Eq_A3}) into Eq. (\ref{Eq_2_5}), we get,
% ------------------
\begin{eqnarray}
\varepsilon = \delta  + \left( - \frac{1}{2} + {\frac{\pi }{4}} \right){\delta ^2} - \frac{\pi }{{32}}{\delta ^4} - \frac{\pi }{{32}}{\delta ^5} +  \ldots, 
\label{Eq_A4}
\end{eqnarray}
% ------------------
% --
\noindent
or the following equivalent expression,
% ------------------
\begin{eqnarray}
\delta  = \varepsilon \left[ {1 + \left( {\frac{1}{2} - \frac{\pi }{4}} \right)\varepsilon  + \frac{\left( \pi - 2  \right)^2}{8} {\varepsilon ^2} + \left( {\frac{5}{8} - \frac{{29\pi }}{{32}} + \frac{{15{\pi ^2}}}{{32}} - \frac{{5{\pi ^3}}}{{64}}} \right){\varepsilon ^3} +  \ldots } \right],
\label{Eq_A5}
\end{eqnarray}
% ------------------
% --
\noindent
in which, $ \varepsilon = (h/2)/ x_{2} $. Substitute Eq. (\ref{Eq_A1}) - (\ref{Eq_A5}) into Eq. (\ref{Eq_2_6}), Eq. (\ref{Eq_2_7}) and Eq. (\ref{Eq_2_14}), we easily obtain Eq. (\ref{Eq_3_1}) - (\ref{Eq_3_5}). Furthermore, on the basis of Eq. (\ref{Eq_3_6}), we give the following dimensionless relationships up to the second order of $\varepsilon$,
% ------------------
\begin{eqnarray}
2 \bar H = 2 + \frac{1}{{{\bar x_2}}}\left[ {\frac{\pi }{4} + \left( \frac{\pi }{4} - \frac{\pi^2 }{16} \right)\varepsilon  +  \left( \frac{9 \pi}{32} - \frac{3 \pi^2}{16} + \frac{\pi^3}{32} \right) \varepsilon^2 + \cdots } \right],
\label{Eq_A6}
\end{eqnarray}
% ------------------
\begin{eqnarray}
\bar U = 2\pi \bar x_2^2\left[ {1 + \left( {\pi  - 2} \right)\varepsilon  + \left( {4 - \frac{\pi }{2} - \frac{{{\pi ^2}}}{4}} \right){\varepsilon ^2} +  \cdots } \right],
\label{Eq_A7}
\end{eqnarray}
% ------------------
\begin{eqnarray}
\bar V = \pi \bar x_2^2\left[ {1 + \left( {\frac{\pi }{2} - 2} \right)\varepsilon  + \left( {\frac{8}{3} - \frac{{{\pi ^2}}}{4}} \right){\varepsilon ^2} +  \cdots } \right],
\label{Eq_A8}
\end{eqnarray}
% ------------------
\begin{eqnarray}
\bar \Delta  = \frac{1}{2} - \frac{1}{2}\left[ {\left( {\frac{\pi }{4} - \frac{1}{2}} \right)\varepsilon  - \frac{1}{2}{{\left( {\frac{\pi }{2} - 1} \right)}^2}{\varepsilon ^2} +  \ldots } \right],
\label{Eq_A9}
\end{eqnarray}
% ------------------
\begin{eqnarray}
\bar F = 2 \pi \bar x_2 \left[ \frac{1}{2 \varepsilon} + \left( -1 + \frac{\pi}{8} \right) + \left( \frac{\pi}{8} - \frac{\pi^2}{32} \right) \varepsilon + \left( \frac{9}{64} \pi -\frac{3}{32} \pi^2 + \frac{1}{64} \pi^3 \right) \varepsilon^2 + \ldots \right].
\label{Eq_A10}
\end{eqnarray}
% --------------------------------------------------------------------------------------------------
\subsection{Wetting states}\label{appA_2}
If $ \theta = 0^{\circ} $, we get $ a = -x_{2}/x_{1} $ and $ k^{2} = 1-a^{2}(x_{2}/x_{1})^{2} $. When $ h \ll x_{2} $, we define $ \Delta = x_{1} - x_{2} $ and $ \delta = \Delta / x_{2} $. Herein, $ a= -1 / (1+\delta) $ and $ k = \sqrt{1-1/(1+\delta)^4} $. Since $\Delta \to h/2$, $ \delta \ll 1 $, we also choose $\delta$ as an infinitesimal to make Taylor series. In this case,
% ------------------
\begin{eqnarray}
\varphi_1 = \frac{\pi}{4} + \frac{1}{2} \delta - \frac{1}{4} \delta^2 + \frac{1}{12} \delta^3 + \ldots,
\label{Eq_A11}
\end{eqnarray}
% ------------------
\begin{eqnarray}
F\left( {{\varphi _1},k} \right) |_{\varphi_1}^{\pi / 2} = \frac{\pi }{4} + \frac{\pi }{4}\delta - \frac{{\pi }}{{16}}{\delta ^2} + \frac{5 \pi }{256}{\delta ^4} + \ldots,
\label{Eq_A12}
\end{eqnarray}
% ------------------
\begin{eqnarray}
E\left( {{\varphi _1},k} \right) |_{\varphi_1}^{\pi / 2} = \frac{\pi }{4} - \left( {1 + \frac{\pi }{4}} \right)\delta  + \left( { \frac{3}{2} + \frac{{7\pi }}{{16}}} \right){\delta ^2} - \left( 2 + \frac{5 \pi}{8} \right) \delta^3+  \ldots 
\label{Eq_A13}
\end{eqnarray}
% ------------------
% --
\noindent
Put Eq. (\ref{Eq_A11}) - (\ref{Eq_A13}) into Eq. (\ref{Eq_2_10}), we get,
% ------------------
\begin{eqnarray}
\varepsilon = \delta  + \left( {\frac{1}{2} - \frac{\pi }{4}} \right){\delta ^2} + \frac{\pi }{{32}}{\delta ^4} - \frac{\pi }{{32}}{\delta ^5} +  \ldots, 
\label{Eq_A14}
\end{eqnarray}
% ------------------
% --
\noindent
or we rewrite Eq. (\ref{Eq_A14}) into,
% ------------------
\begin{eqnarray}
\delta  = \varepsilon \left[ {1 - \left( {\frac{1}{2} - \frac{\pi }{4}} \right)\varepsilon  + \frac{\left( \pi - 2  \right)^2}{8} {\varepsilon ^2} - \left( { \frac{5}{8} - \frac{{29\pi }}{{32}} + \frac{{15{\pi ^2}}}{{32}} - \frac{{5{\pi ^3}}}{{64}}} \right){\varepsilon ^3} +  \ldots } \right],
\label{Eq_A15}
\end{eqnarray}
% ------------------
% --
\noindent
in which, $ \varepsilon = (h/2)/ x_{2} $. Put Eq. (\ref{Eq_A11}) - (\ref{Eq_A15}) into Eq. (\ref{Eq_2_11}), Eq. (\ref{Eq_2_12}) and Eq. (\ref{Eq_2_16}), we finally get Eq. (\ref{Eq_3_7}) - (\ref{Eq_3_11}). Furthermore, the resulting dimensionless relationships are obtained using Eq. (\ref{Eq_3_6}),
% ------------------
\begin{eqnarray}
2 \bar H = -2 + \frac{1}{{{\bar x_2}}}\left[ {\frac{\pi }{4} - \left( \frac{\pi }{4} - \frac{\pi^2 }{16} \right)\varepsilon  +  \left( \frac{9 \pi}{32} - \frac{3 \pi^2}{16} + \frac{\pi^3}{32} \right) \varepsilon^2 + \cdots } \right],
\label{Eq_A16}
\end{eqnarray}
% ------------------
\begin{eqnarray}
\bar U = 2\pi \bar x_2^2\left[ { - 1 + \left( {\pi  - 2} \right)\varepsilon  + \left( { - 4 + \frac{\pi }{2} + \frac{{{\pi ^2}}}{4}} \right){\varepsilon ^2} +  \cdots } \right],
\label{Eq_A17}
\end{eqnarray}
% ------------------
\begin{eqnarray}
\bar V = \pi \bar x_2^2\left[ {1 + \left( { - \frac{\pi }{2} + 2} \right)\varepsilon  + \left( {\frac{8}{3} - \frac{{{\pi ^2}}}{4}} \right){\varepsilon ^2} +  \cdots } \right],
\label{Eq_A18}
\end{eqnarray}
% ------------------
\begin{eqnarray}
\bar \Delta  = \frac{1}{2} + \frac{1}{2}\left[ {\left( { \frac{\pi}{4} - \frac{1}{2}} \right)\varepsilon  + \frac{1}{2}{{\left( {\frac{\pi }{2} - 1} \right)}^2}{\varepsilon ^2} +  \ldots } \right],
\label{Eq_A19}
\end{eqnarray}
% ------------------
\begin{eqnarray}
\bar F = 2 \pi \bar x_2 \left[ - \frac{1}{2 \varepsilon} + \left( -1 + \frac{\pi}{8} \right) - \left( \frac{\pi}{8} - \frac{\pi^2}{32} \right) \varepsilon + \left( \frac{9}{64} \pi -\frac{3}{32} \pi^2 + \frac{1}{64} \pi^3 \right) \varepsilon^2 + \ldots \right].
\label{Eq_A20}
\end{eqnarray}
% ------------------
% --------------------------------------------------------------------------------------------------
\section{Liquid Hertzian contact}\label{appB}
% --
When $\theta = 180^{\circ}$ and $\sigma = x_1 / x_2 \ll 1$, we get $a=-\sigma$ and $ k = \sqrt{1 - \sigma^4} $. Putting them into the boundary conditions (i.e. $\varphi_1 = \arcsin \sqrt{1/(1+\sigma^2)}$) of Eq. (\ref{Eq_2_1}) - Eq. (\ref{Eq_2_3}) leads to, 
% ------------------
\begin{eqnarray}
\varphi_1 = \frac{\pi}{2} - \sigma + \frac{1}{3} \sigma^3 - \frac{1}{5} \sigma^5 + \ldots,
\label{Eq_B1}
\end{eqnarray}
% ------------------
% --
\noindent
However, in this case, it is not so straightforward to expand the elliptic integrals into power series. The following transformation is employed for our further calculations, 
% ------------------
\begin{eqnarray}
\nonumber
\sqrt{1 - k^2 \sin^2 \varphi} & = & \cos \varphi \sqrt{ 1 + \sigma^4 \tan^2 \varphi } \\
& = & \cos \varphi \left(  1 + \frac{1}{2} \sigma^4 \tan^2 \varphi - \frac{1}{8} \sigma^8 \tan^4 \varphi + \frac{1}{16} \sigma^{12} \tan^6 \varphi + \cdots \right)
\label{Eq_B2}
\end{eqnarray}
% ------------------
% --
\noindent
Subsequently, we can get the following expressions based on Eq (\ref{Eq_B2}),
% ------------------
\begin{eqnarray}
F\left( {{\varphi _1},k} \right) = - \ln {\frac{\sigma}{2}} - \frac{\sigma^4}{4} \ln \left( \frac{e^{1/2}}{2} \sigma \right) - \frac{9 \sigma^8}{64} \ln \left( \frac{e^{7/12}}{2} \sigma \right) + \ldots 
\label{Eq_B3}
\end{eqnarray}
% ------------------
\begin{eqnarray}
E\left( {{\varphi _1},k} \right) = 1 - \frac{\sigma^2}{2} - \frac{\sigma^4}{2} \ln \left( \frac{e^{1/4}}{2} \sigma \right) - \frac{3 \sigma^8}{16} \ln \left( \frac{e^{13/24}}{2} \sigma \right) + \ldots 
\label{Eq_B4}
\end{eqnarray}
% ------------------
% --------------------------------------------------------------------------------------------------
\subsection{Constant $h$}\label{appB_1}
% ************
\begin{figure}
  \centerline{\includegraphics[width=11.0cm]{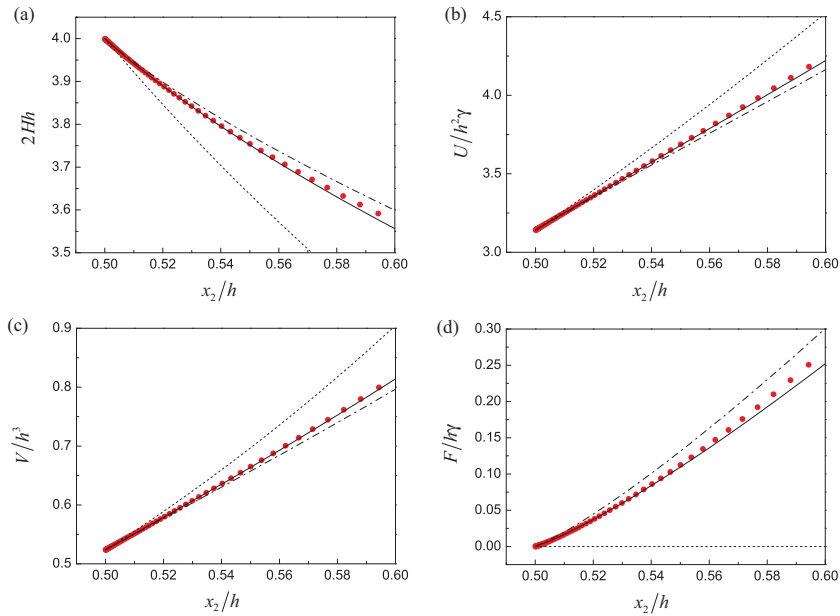}}
  \caption{Dependence of the curvature $2H$, surface energy $U$, volume $V$ and capillary force $F$ of the liquid as functions of $x_2$ in dimensionless form. The red dots in each figure are numerical results based on the analytical solutions (Eq.(\ref{Eq_2_5}) - (\ref{Eq_2_7}), Eq.(\ref{Eq_2_14})). The dotted lines represent $ 2H \approx 2/x_2 $, $ U \approx 4 \pi x_2^2 \gamma $, $ V \approx 4 \pi x_2^3 /3 $ and $ F \approx 0 $, respectively. The dashed lines represent that we consider the term up to $\xi \left[ - \ln \left( \frac{1}{2e^{1/2}} \xi^{1/2} \right) \right]^{-1} $ in Eq. (\ref{Eq_B7}) - (\ref{Eq_B10}). And the solid lines mean the higer order of $\xi$ are included (i.e., up to the expressions of Eq. (\ref{Eq_B7}) - (\ref{Eq_B10}) we give).}
\label{fig:figS1}
\end{figure}
% ************
Since $ h $ is fixed, substituting Eq. (\ref{Eq_B1}), (\ref{Eq_B3}) and (\ref{Eq_B4}) into Eq. (\ref{Eq_2_5}) leads to,
% ------------------
\begin{eqnarray}
2(1 - \xi) = 2 + 2 \sigma ^{2} \ln \left( \frac{1}{2e^{1/2}} \sigma \right) - \sigma^4 \ln \left( \frac{e^{1/4}}{2} \sigma \right) + \frac{\sigma^6}{2} \ln \left( \frac{e^{1/2}}{2} \sigma \right) \cdots
\label{Eq_B5}
\end{eqnarray}
% ------------------
% --
\noindent
accordingly, we can get Eq. (\ref{Eq_4_2}), or equivalently, we express $\sigma$ as the function of $\xi$ (i.e., $\bar x_2$) using some asymptotic method,
% ------------------
\begin{eqnarray}
\sigma = \xi^{1/2} \cdot \left[ - \ln \left( \frac{1}{2e^{1/2}} \xi^{1/2} \right) \right]^{ - \frac{1}{2} + \frac{1}{4} \ln^{-1} (\frac{1}{2e^{1/2}} \xi^{1/2}) + \cdots }
\label{Eq_B6}
\end{eqnarray}
% ------------------
% --
\noindent
in which $ \xi = 1- h/(2x_2) = 1 - 1/(2 \bar x_2) $. It should be noted that this procedure can be extended to higher orders of $\xi$ by employing the more elaborate mathematical techniques, but which is beyond the scope of this paper. Furthermore, we can get the following relationships as functions of $h$ and $x_2$ only,
% ------------------
\begin{eqnarray}
\nonumber
2H & = & \frac{2}{x_2} (1 + \sigma^2 + \sigma^4 + \cdots ) \\
& = & \frac{2}{x_2} \left\{ 1+ \xi \left[ -\ln \left( \frac{1}{2e^{1/2}} \xi^{1/2} \right) \right]^{ -1 + \frac{1}{2} \ln^{-1} \left( \frac{1}{2e^{1/2}} \xi^{1/2} \right) + \cdots } \right\},
\label{Eq_B7}
\end{eqnarray}
% ------------------
\begin{eqnarray}
\nonumber
U & = & 4 \pi x_2^2 \gamma \left[ 1 - \sigma^2 - \frac{1}{2} \sigma^4 \ln \left( \frac{1}{2e^{3/4}} \sigma \right) + \cdots \right] \\
& = & 4\pi x_2^{2} \gamma \left\{ 1+ \xi \left[ -\ln \left( \frac{1}{2e^{1/2}} \xi^{1/2} \right) \right]^{ -1 + \frac{1}{2} \ln^{-1} \left( \frac{1}{2e^{1/2}} \xi^{1/2} \right) + \cdots } \right\},
\label{Eq_B8}
\end{eqnarray}
% ------------------
\begin{eqnarray}
\nonumber
V & = & \frac{4 \pi}{3} x_2^3 (1 - \frac{3}{2} \sigma^2 + \frac{9}{8} \sigma^4 + \cdots) \\
& = & \frac{4\pi }{3} x_2^3 \left\{ 1- \frac{3\xi}{2} \left[ -\ln \left( \frac{1}{2e^{1/2}} \xi^{1/2} \right) \right]^{ -1 + \frac{1}{2} \ln^{-1} \left( \frac{1}{2e^{1/2}} \xi^{1/2} \right) + \cdots } \right\},
\label{Eq_B9}
\end{eqnarray}
% ------------------
\begin{eqnarray}
\nonumber
F & = & 2 \pi x_2 \gamma \left( \sigma^2 + \sigma^4 + \cdots \right) \\
& = & 2 \pi x_2 \gamma \xi \left[ -\ln \left( \frac{1}{2e^{1/2}} \xi^{1/2} \right) \right]^{ -1 + \frac{1}{2} \ln^{-1} \left( \frac{1}{2e^{1/2}} \xi^{1/2} \right) + \cdots },
\label{Eq_B10}
\end{eqnarray}
% ------------------
\begin{eqnarray}
\nonumber
\Delta & = & x_2 \left( 1 - \sigma \right) \\
& = & x_2 \left\{ 1 - \xi^{1/2} \cdot \left[ - \ln \left( \frac{1}{2e^{1/2}} \xi^{1/2} \right) \right]^{ - \frac{1}{2} + \frac{1}{4} \ln^{-1} (\frac{1}{2e^{1/2}} \xi^{1/2}) + \cdots } \right\}.
\label{Eq_B11}
\end{eqnarray}
% ------------------

% --
In figure \ref{fig:figS1}, comparisons between the numerical results (the red dots) and approximate solutions up to different orders of $\xi$ in Eq. (\ref{Eq_B7}) - (\ref{Eq_B10}) are given. The dotted lines represent $ 2H \approx 2/x_2 $, $ U \approx 4 \pi x_2^2 \gamma $, $ V \approx 4 \pi x_2^3 /3 $ and $ F \approx 0 $. The dashed lines represent that we consider the term up to $\xi \left[ - \ln \left( \frac{1}{2e^{1/2}} \xi^{1/2} \right) \right]^{-1} $ in Eq. (\ref{Eq_B7}) - (\ref{Eq_B10}). And the solid lines means that the higher order of $\xi$ are included (i.e., up to the expressions of Eq. (\ref{Eq_B7}) - (\ref{Eq_B10}) we give).
% --------------------------------------------------------------------------------------------------
\subsection{Constant $V$}\label{appB_2}
% ************
\begin{figure}
  \centerline{\includegraphics[width=11.0cm]{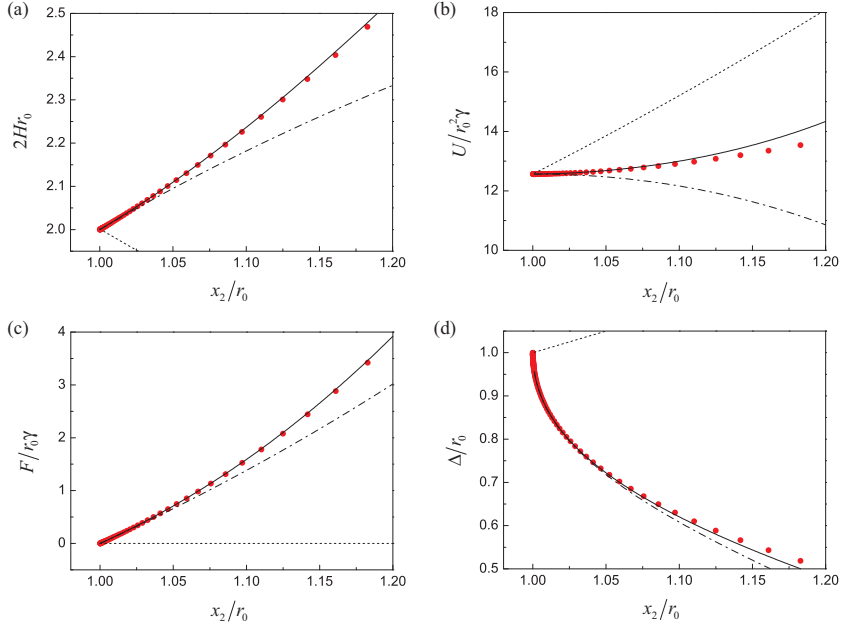}}
  \caption{Dependence of the curvature $2H$, $U$, $F$ and $ \Delta $ as functions of $x_2$ in dimensionless form. The red dots in each figure are the numerical results of the analytical solutions (Eq.(\ref{Eq_2_10}) - (\ref{Eq_2_12}), Eq.(\ref{Eq_2_16})). ({\it a})({\it d})({\it c})  The dotted/dashed/solid lines means we keep $(\hat x_2 -1)$ up to the zero/first/second order in Eq. (\ref{Eq_B19}) - (\ref{Eq_B21}), respectively. ({\it d}) The dotted, dashed, solid lines correspond to that we keep $(\hat x_2 -1)$ up to the zero, 1/2 and 3/2 orders in Eq. (\ref{Eq_B22}), respectively.}
\label{fig:figS2}
\end{figure}
% ************
% --
Since $ V $ is fixed, substituting Eq. (\ref{Eq_B1}), (\ref{Eq_B3}) and (\ref{Eq_B4}) into Eq. (\ref{Eq_2_7}), a volume conservation constraint, leads to,
% ------------------
\begin{eqnarray}
V = \frac{4 \pi}{3} r_0^3 = \frac{4 \pi}{3} x_2^3 \left[ 1 - \frac{3}{2} \sigma^2 + \frac{9}{8} \sigma^4 + \frac{3}{4} \sigma^6 \ln \left( \frac{1}{2e^{5/12}} \sigma \right) + \cdots \right]
\label{Eq_B12}
\end{eqnarray}
% ------------------
% --
\noindent
and accordingly, 
% ------------------
\begin{eqnarray}
\hat x_2 - 1 = \frac{1}{2} \sigma^2 + \frac{1}{8} \sigma^4 - \frac{1}{4} \ln \left( \frac{e^{1/4}}{2} \sigma \right) + \cdots
\label{Eq_B13}
\end{eqnarray}
% ------------------
% --
\noindent
Here, $ (\hat x_2 - 1) \ll 1 $. Considering $ \sigma = x_1/ x_2 = \hat x_1 / \hat x_2 $, we can get,
% ------------------
\begin{eqnarray}
\hat x_1 = \sqrt{2} \left( \hat x_2 - 1 \right)^{1/2} + \frac{3\sqrt{2}}{4} \left( \hat x_2 - 1 \right)^{3/2} - \frac{\sqrt{2}}{4} \left( \hat x_2 - 1 \right)^{5/2} + \cdots
\label{Eq_B14}
\end{eqnarray}
% ------------------
\begin{eqnarray}
\hat x_1^2 = 2 \left( \hat x_2 - 1 \right) + 3 \left( \hat x_2 - 1 \right)^2 + \frac{1}{8} \left( \hat x_2 - 1 \right)^3 + \cdots
\label{Eq_B15}
\end{eqnarray}
% ------------------
\begin{eqnarray}
\nonumber
h & = & 2 x_2 \left[ 1 + \sigma^2 \ln \left( \frac{1}{2e^{1/2}} \sigma \right) - \frac{\sigma^4}{2} \ln \left( \frac{e^{1/4}}{2} \sigma \right) +  \cdots \right] \\
& = & 2 x_2 \left[ 1 + \left( \hat x_2 - 1 \right) \ln \left( \frac{ \hat x_2 - 1 }{2e} \right) - \frac{3}{2} \left( \hat x_2 - 1 \right)^2 \ln \left( \frac{\hat x_2 - 1}{2 e^{-1/3}} \right) + \cdots \right] 
\label{Eq_B16}
\end{eqnarray}
% ------------------
% --
Let's define $ \Omega = r_0 - h/2 $, and we make an analysis about the scaling law obeyed between $ x_1^2 $ and $ \Omega r_0 $. On the basis of Eq. (\ref{Eq_B15}) and Eq. (\ref{Eq_B16}), if we ignore the second and higher orders of $ \left( \hat x_2 - 1 \right) $, the following relationships can be derived,
% ------------------
\begin{eqnarray}
\ln \left( \Omega r_0 \right) = 2 r_0 + \ln \left( \hat x_2 - 1 \right) + \ln \left[ - \ln \left( \frac{\hat x_2 - 1}{2} \right) \right] + \cdots
\label{Eq_B17}
\end{eqnarray}
% ------------------
\begin{eqnarray}
\ln \left( x_1^2 \right) = 2 r_0 + \ln 2 + \ln \left( \hat x_2 - 1 \right) + \cdots
\label{Eq_B18}
\end{eqnarray}
% ------------------
% --
\noindent
Since $ \ln (\hat x_2 - 1) $ is dominant in Eq. (\ref{Eq_B17}) and (\ref{Eq_B18}), so $ \ln (\Omega r_0) \sim \ln (\Omega x_2) \sim \ln(x_1^2) $, and subsequently $ \Omega r_0  \sim \Omega x_2 \sim x_1^2 $.
% --
Next, similar to the last section, we can get the following relationships,
% ------------------
\begin{eqnarray}
\nonumber
2H & = & \frac{2}{x_2} (1 + \sigma^2 + \sigma^4 + \cdots ) \\
& = & \frac{2}{x_2} \left[ 1+ 2 \left( \hat x_2 - 1 \right) + 3 \left( \hat x_2 - 1 \right)^2 + \cdots \right],
\label{Eq_B19}
\end{eqnarray}
% ------------------
\begin{eqnarray}
\nonumber
U & = & 4 \pi x_2^2 \gamma \left[ 1 - \sigma^2 - \frac{1}{2} \sigma^4 \ln \left( \frac{1}{2e^{3/4}} \sigma \right) + \cdots \right] \\
& = & 4\pi x_2^2 \gamma \left[ 1 - 2 \left( \hat x_2 - 1 \right) - \left( \hat x_2 - 1 \right)^2 \ln \left( \frac{\hat x_2 - 1}{2e^{5/2}} \right) + \cdots \right],
\label{Eq_B20}
\end{eqnarray}
% ------------------
\begin{eqnarray}
\nonumber
F & = & 2 \pi x_2 \gamma \left( \sigma^2 + \sigma^4 + \cdots \right) \\
& = & 4 \pi x_2 \gamma \left[ \left( \hat x_2 - 1 \right) + \frac{3}{2} \left( \hat x_2 - 1 \right)^2 + \cdots \right]
\label{Eq_B21}
\end{eqnarray}
% ------------------
\begin{eqnarray}
\nonumber
\Delta & = & x_2 \left( 1 - \sigma \right) \\
& = & x_2 \left[ 1 - \sqrt{2} \left( \hat x_2 - 1 \right)^{1/2} + \frac{\sqrt 2}{4} \left( \hat x_2 - 1 \right)^{3/2} + \cdots \right] 
\label{Eq_B22}
\end{eqnarray}
% ------------------
% --
\noindent
In order to check the availability of these asymptotic solutions, we compare them with the analytical results solved numerically (the red dots) in figure \ref{fig:figS2}. In figure \ref{fig:figS2}(a)(b)(c), the dotted/dashed/solid lines means that we keep $(\hat x_2 - 1)$ up to the zero/first/second orders in Eq. (\ref{Eq_B19}) - (\ref{Eq_B21}), respectively. In figure \ref{fig:figS2}(d), the dotted, dashed and solid lines correspond to that we keep the zero, $1/2$ and $3/2$ orders of $(\hat x_2 - 1)$ in Eq. (\ref{Eq_B22}), respectively.
% --------------------------------------------------------------------------------------------------
\bibliographystyle{jfm}
\bibliography{Sandwich_droplet}

\begin{thebibliography}{49}
\expandafter\ifx\csname natexlab\endcsname\relax\def\natexlab#1{#1}\fi
\def\au#1{#1} \def\ed#1{#1} \def\yr#1{#1}\def\at#1{#1}\def\jt#1{\textit{#1}}
  \def\bt#1{#1}\def\bvol#1{\textbf{#1}} \def\vol#1{#1} \def\pg#1{#1}
  \def\publ#1{#1}\def\arxiv#1{#1}\def\org#1{#1}\def\st#1{\textit{#1}}

\bibitem[Ajaev \& Homsy(2006)]{Ajaev2006}
{\sc \au{Ajaev, V.~S.} \& \au{Homsy, G.~M.}} \yr{2006}  \at{Modeling shapes and
  dynamics of confined bubbles}.  \jt{Annu. Rev. Fluid Mech.}  \bvol{38},
  \pg{277--307}.

\bibitem[Amirfazli \& Neumann(2004)]{Amirfazli2004}
{\sc \au{Amirfazli, A.} \& \au{Neumann, A.~W.}} \yr{2004}  \at{Status of the
  three-phase line tension}.  \jt{Adv. Colloid Interface Sci.}  \bvol{110},
  \pg{121--141}.

\bibitem[Amselem {\em et~al.\/}(2015)Amselem, Brun, Gallaire \&
  Baroud]{Amselem2015}
{\sc \au{Amselem, G.}, \au{Brun, P.~T.}, \au{Gallaire, F.} \& \au{Baroud,
  C.~N.}} \yr{2015}  \at{Breaking anchored droplets in a microfluidic hele-shaw
  cell}.  \jt{Phys. Rev. Applied}  \bvol{3},  \pg{054006}.

\bibitem[Aussillous \& Qu{\'e}r{\'e}(2001)]{Aussillous2001}
{\sc \au{Aussillous, P.} \& \au{Qu{\'e}r{\'e}, D.}} \yr{2001}  \at{Liquid
  marbles}.  \jt{Nature}  \bvol{411},  \pg{924--927}.

\bibitem[Bensimon {\em et~al.\/}(1986)Bensimon, Kadanoff, Liang, Shraiman \&
  Tang]{Bensimon1986}
{\sc \au{Bensimon, D.}, \au{Kadanoff, L.~P.}, \au{Liang, S.}, \au{Shraiman,
  B.~I.} \& \au{Tang, C.}} \yr{1986}  \at{Viscous flows in two dimensions}.
  \jt{Rev. Mod. Phys.}  \bvol{58},  \pg{977--999}.

\bibitem[Bonn {\em et~al.\/}(2009)Bonn, Eggers, Indekeu, Meunier \&
  Rolley]{Bonn2009}
{\sc \au{Bonn, D.}, \au{Eggers, J.}, \au{Indekeu, J.}, \au{Meunier, J.} \&
  \au{Rolley, E.}} \yr{2009}  \at{Wetting and spreading}.  \jt{Rev. Mod. Phys.}
   \bvol{81},  \pg{739--805}.

\bibitem[Bostwick \& Steen(2015)]{Bostwick2015}
{\sc \au{Bostwick, J.~B.} \& \au{Steen, P.~H.}} \yr{2015}  \at{Stability of
  constrained capillary surfaces}.  \jt{Annu. Rev. Fluid Mech.}  \bvol{47},
  \pg{539--568}.

\bibitem[Butt \& Kappl(2009)]{Butt2009}
{\sc \au{Butt, H.-J.} \& \au{Kappl, M.}} \yr{2009}  \at{Normal capillary
  forces}.  \jt{Adv. Colloid Interf. Sci.}  \bvol{146},  \pg{48--60}.

\bibitem[Carroll(1976)]{Carroll1976}
{\sc \au{Carroll, B.~J.}} \yr{1976}  \at{The accurate measurement of contact
  angle, phase contact areas, drop volume, and laplace excess pressure in
  drop-on-fiber systems}.  \jt{J. Colloid Interf. Sci.}  \bvol{57},
  \pg{488--495}.

\bibitem[Chevy {\em et~al.\/}(2012)Chevy, Chepelianskii, Qu{\'e}r{\'e} \&
  Rapha{\"e}l]{Chevy2012}
{\sc \au{Chevy, F.}, \au{Chepelianskii, A.}, \au{Qu{\'e}r{\'e}, D.} \&
  \au{Rapha{\"e}l, E.}} \yr{2012}  \at{Liquid hertz contact: softness of weakly
  deformed drops on non-wetting substrates}.  \jt{Euro. Phys. Lett.}
  \bvol{100},  \pg{54002}.

\bibitem[Christenson \& Claesson(1988)]{Christenson1988}
{\sc \au{Christenson, H.~K.} \& \au{Claesson, P.~M.}} \yr{1988}  \at{Cavitation
  and the interaction between macroscopic hydrophobic surfaces}.  \jt{Science}
  \bvol{239},  \pg{390--392}.

\bibitem[Dangla {\em et~al.\/}(2013)Dangla, Kayi \& Baroud]{Dangla2013}
{\sc \au{Dangla, R.}, \au{Kayi, S.~C.} \& \au{Baroud, C.~N.}} \yr{2013}
  \at{Droplet microfluidics driven by gradients of confinement}.  \jt{PNAS}
  \bvol{110},  \pg{853--858}.

\bibitem[Dangla {\em et~al.\/}(2011)Dangla, LeeB. \& Baroud]{Dangla2011}
{\sc \au{Dangla, R.}, \au{LeeB., S.} \& \au{Baroud, C.~N.}} \yr{2011}
  \at{Trapping microfluidic drops in wells of surface energy}.  \jt{Phys. Rev.
  Lett.}  \bvol{107},  \pg{124501}.

\bibitem[Eddi {\em et~al.\/}(2013)Eddi, Winkels \& Snoeijer]{Eddi2013}
{\sc \au{Eddi, A.}, \au{Winkels, K.~G.} \& \au{Snoeijer, J.~H.}} \yr{2013}
  \at{Influence of droplet geometry on the coalescence of low viscosity drops}.
   \jt{Phys. Rev. Lett.}  \bvol{111},  \pg{144502}.

\bibitem[Eichelsdoerfer {\em et~al.\/}(2014)Eichelsdoerfer, Brown \&
  Mirkin]{Eichelsdoerfer2014}
{\sc \au{Eichelsdoerfer, D.~J.}, \au{Brown, K.~A.} \& \au{Mirkin, C.~A.}}
  \yr{2014}  \at{Capillary bridge rupture in dip-pen nanolithography}.
  \jt{Soft Matter}  \bvol{10},  \pg{5603--5608}.

\bibitem[Federle {\em et~al.\/}(2002)Federle, Riehle, Curtis \&
  Full]{Federle2002}
{\sc \au{Federle, W.}, \au{Riehle, M.}, \au{Curtis, A. S.~G.} \& \au{Full,
  R.~J.}} \yr{2002}  \at{An integrative study of insect adhesion: mechanics and
  wet adhesion of pretarsal pads in ants}.  \jt{Integr. Comp. Biol.}
  \bvol{42},  \pg{1100--1106}.

\bibitem[Fisher(1926)]{Fisher1926}
{\sc \au{Fisher, R.~A.}} \yr{1926}  \at{On the capillary forces in an ideal
  soil}.  \jt{Integr. Comp. Biol.}  \bvol{16},  \pg{492--505}.

\bibitem[de~Gennes(1985)]{deGennes1985}
{\sc \au{de~Gennes, P.~G.}} \yr{1985}  \at{Wetting: statics and dynamics}.
  \jt{Rev. Mod. Phys.}  \bvol{57},  \pg{827--863}.

\bibitem[de~Gennes {\em et~al.\/}(2004)de~Gennes, Brochard-Wyart \&
  Qu{\'e}r{\'e}]{deGennes2004}
{\sc \au{de~Gennes, P.-G.}, \au{Brochard-Wyart, F.} \& \au{Qu{\'e}r{\'e}, D.}}
  \yr{2004}  \bt{ \at{{\it Capillarity and Wetting Phenomena: Drops, Bubbles,
  Pearls and Waves}}}.  \publ{{\rm Springer}}.

\bibitem[Geoffroy {\em et~al.\/}(2006)Geoffroy, Plourabou{\'e}, Prat \&
  Amyot]{Geoffroy2006}
{\sc \au{Geoffroy, S.}, \au{Plourabou{\'e}, F.}, \au{Prat, M.} \& \au{Amyot,
  O.}} \yr{2006}  \at{Quasi-static liquid-air drainage in narrow channels with
  variations in the gap}.  \jt{J. Colloid Interface Sci}  \bvol{294},
  \pg{165--175}.

\bibitem[Gondret \& Rabaud(1997)]{Gondret1997}
{\sc \au{Gondret, P.} \& \au{Rabaud, M.}} \yr{1997}  \at{Shear instability of
  two-fluid parallel flow in a hele-shaw cell}.  \jt{Phys. Fluids}  \bvol{9},
  \pg{3267--3274}.

\bibitem[Haines(1925)]{Haines1925}
{\sc \au{Haines, W.~B.}} \yr{1925}  \at{Studies in the physical properties of
  soils. ii. a note on the cohesion developed by capillary forces in an ideal
  soil}.  \jt{J. Agric. Sci.}  \bvol{15},  \pg{529--535}.

\bibitem[Israelachvili(2004)]{Israelachvili2011}
{\sc \au{Israelachvili, J.~N.}} \yr{2004}  \bt{ \at{{\it Intermolecular and
  Surface Forces (Third Edition)}}}.  \publ{{\rm Elsevier Science}}.

\bibitem[Johnson(1986)]{Johnson1986}
{\sc \au{Johnson, K.~L.}} \yr{1986}  \bt{ \at{{\it Contact Mechanics}}}.
  \publ{{\rm Cambridge University Press, Cambridge}}.

\bibitem[Kavahpour(2015)]{Kavahpour2015}
{\sc \au{Kavahpour, H.~P.}} \yr{2015}  \at{Coalescence of drops}.  \jt{Annu.
  Rev. Fluid Mech.}  \bvol{47},  \pg{245--268}.

\bibitem[Keiser {\em et~al.\/}(2016)Keiser, Herbaut, Bico \&
  Reyssat]{Keiser2016}
{\sc \au{Keiser, L.}, \au{Herbaut, R.}, \au{Bico, J.} \& \au{Reyssat, E.}}
  \yr{2016}  \at{Washing wedges: capillary instability in a gradient of
  confinement}.  \jt{J. Fluid Mech.}  \bvol{790},  \pg{619--633}.

\bibitem[Lafuma \& Qu{\'e}r{\'e}(2003)]{Lafuma2003}
{\sc \au{Lafuma, A.} \& \au{Qu{\'e}r{\'e}, D.}} \yr{2003}  \at{Superhydrophobic
  states}.  \jt{Nat. Mater.}  \bvol{2},  \pg{457--460}.

\bibitem[Lai {\em et~al.\/}(2009)Lai, Bremond \& Stone]{Lai2009}
{\sc \au{Lai, A.}, \au{Bremond, N.} \& \au{Stone, H.~A.}} \yr{2009}
  \at{Separation-driven coalescence of droplets: an analytical criterion for
  the approach to contact}.  \jt{J. Fluid Mech.}  \bvol{632},  \pg{97--107}.

\bibitem[Lv {\em et~al.\/}(2015)Lv, Hao, Zhang \& He]{Lv2015}
{\sc \au{Lv, C.}, \au{Hao, P.}, \au{Zhang, X.} \& \au{He, F.}} \yr{2015}
  \at{Dewetting transitions of dropwise condensation on nanotexture-enhanced
  superhydrophobic surfaces}.  \jt{ACS Nano}  \bvol{9},  \pg{12311--12319}.

\bibitem[Macner \& Steen(2014)]{Macner2014}
{\sc \au{Macner, A.~M.} \& \au{Steen, P.~H.}} \yr{2014}  \at{Adaptive adhesion
  by a beetle: Manipulation of liquid bridges and their breaking limits}.
  \jt{Biointerphases}  \bvol{9},  \pg{011001}.

\bibitem[Mahadevan \& Pomeau(1999)]{Mahadevan1999}
{\sc \au{Mahadevan, L.} \& \au{Pomeau, Y.}} \yr{1999}  \at{Rolling droplets}.
  \jt{Phys. Fluids}  \bvol{11},  \pg{2449--2453}.

\bibitem[Mitarai \& Nori(2006)]{Mitarai2006}
{\sc \au{Mitarai, N.} \& \au{Nori, F.}} \yr{2006}  \at{Wet granular materials}.
   \jt{Adv. Phys.}  \bvol{55},  \pg{1--45}.

\bibitem[Mol{\'a}{\v{c}}ek \& Bush(2012)]{Molacek2012}
{\sc \au{Mol{\'a}{\v{c}}ek, J.} \& \au{Bush, J. W.~M.}} \yr{2012}  \at{A
  quasi-static model of drop impact}.  \jt{Phys. Fluids}  \bvol{24},
  \pg{127103}.

\bibitem[Orr {\em et~al.\/}(1975)Orr, Scriven \& Rivas]{Orr1975}
{\sc \au{Orr, F.~M.}, \au{Scriven, L.~E.} \& \au{Rivas, A.~P.}} \yr{1975}
  \at{Pendular rings between solids: meniscus properties and capillary force}.
  \jt{J. Fluid Mech.}  \bvol{67},  \pg{723--742}.

\bibitem[Park \& Homsy(1984)]{Park1984}
{\sc \au{Park, C.-W.} \& \au{Homsy, G.~M.}} \yr{1984}  \at{Two-phase
  displacement in hele shaw cells: theory}.  \jt{J. Fluid Mech.}  \bvol{139},
  \pg{291--308}.

\bibitem[Paterson {\em et~al.\/}(1995)Paterson, Fermigier, Jenffer \&
  Limat]{Paterson1995}
{\sc \au{Paterson, A.}, \au{Fermigier, M.}, \au{Jenffer, P.} \& \au{Limat, L.}}
  \yr{1995}  \at{Wetting on heterogeneous surfaces: experiments in an imperfect
  hele-shaw cell}.  \jt{Phys. Rev. E}  \bvol{51},  \pg{1291--1298}.

\bibitem[Peisker \& Gorb(2012)]{Peisker2012}
{\sc \au{Peisker, H.} \& \au{Gorb, S.~N.}} \yr{2012}  \at{Evaporation dynamics
  of tarsal liquid footprints in flies ({\it calliphora vicina}) and ({\it
  coccinella septempunctata}) beetles}.  \jt{J. Exp. Biol.}  \bvol{215},
  \pg{1266--1271}.

\bibitem[Plourabou{\'e} \& Hinch(2002)]{Plouraboue2002}
{\sc \au{Plourabou{\'e}, F.} \& \au{Hinch, E.~J.}} \yr{2002}
  \at{Kelvin-helmholtz instability in a hele-shaw cell}.  \jt{Phys. Fluids}
  \bvol{14},  \pg{922--929}.

\bibitem[Reis {\em et~al.\/}(2010)Reis, Jung, Aristoff \& Stocker]{Reis2010}
{\sc \au{Reis, P.~M.}, \au{Jung, S.}, \au{Aristoff, J.~M.} \& \au{Stocker, R.}}
  \yr{2010}  \at{How cats lap: water uptake by felis catus}.  \jt{Science}
  \bvol{330},  \pg{1231--1234}.

\bibitem[Reyssat(2014)]{Reyssat2014}
{\sc \au{Reyssat, E.}} \yr{2014}  \at{Drops and bubbles in wedges}.  \jt{J.
  Fluid Mech.}  \bvol{748},  \pg{641--662}.

\bibitem[Reyssat(2015)]{Reyssat2015}
{\sc \au{Reyssat, E.}} \yr{2015}  \at{Capillary bridges between a plane and a
  cylindrical wall}.  \jt{J. Fluid Mech.}  \bvol{773},  \pg{773R1}.

\bibitem[Riedo {\em et~al.\/}(2002)Riedo, L{\'e}vy \& Brune]{Riedo2002}
{\sc \au{Riedo, E.}, \au{L{\'e}vy, F.} \& \au{Brune, H.}} \yr{2002}
  \at{Kinetics of capillary condensation in nanoscopic sliding friction}.
  \jt{Phys. Rev. Lett.}  \bvol{889},  \pg{185505}.

\bibitem[Roman \& Bico(2010)]{Roman2010}
{\sc \au{Roman, B.} \& \au{Bico, J.}} \yr{2010}  \at{Elasto-capillarity:
  deforming an elastic structure with a liquid droplet}.  \jt{J. Phys.:
  Condens. Matter}  \bvol{22},  \pg{493101}.

\bibitem[Snoeijer {\em et~al.\/}(2013)Snoeijer, Eggers \& Venner]{Snoeijer2013}
{\sc \au{Snoeijer, J.~H.}, \au{Eggers, J.} \& \au{Venner, C.~H.}} \yr{2013}
  \at{Similarity theory of lubricated hertzian contacts}.  \jt{Phys. Fluids}
  \bvol{25},  \pg{101705}.

\bibitem[Srinivasan {\em et~al.\/}(2011)Srinivasan, McKinley \&
  Cohen]{Srinivasan2011}
{\sc \au{Srinivasan, S.}, \au{McKinley, G.~H.} \& \au{Cohen, R.~E.}} \yr{2011}
  \at{Assessing the accuracy of contact angle measurements for sessile drops on
  liquid-repellent surfaces}.  \jt{Langmuir}  \bvol{27},  \pg{13582--13589}.

\bibitem[Struik(1961)]{Struik1961}
{\sc \au{Struik, D.~J.}} \yr{1961}  \bt{ \at{{\it Lectures on Classical
  Differential Geometry}}}.  \publ{{\rm Addison-Wesley Pub. Co., Dover, New
  York}}.

\bibitem[Thoroddsen {\em et~al.\/}(2005)Thoroddsen, Etoh, Takehara \&
  Ootsuka]{Thoroddsen2005}
{\sc \au{Thoroddsen, S.~T.}, \au{Etoh, T.~G.}, \au{Takehara, K.} \&
  \au{Ootsuka, N.}} \yr{2005}  \at{On the coalescence speed of bubbles}.
  \jt{Phys. Fluids}  \bvol{17},  \pg{071703}.

\bibitem[Wang {\em et~al.\/}(2015)Wang, Yao, Liu, Qu{\'e}r{\'e} \&
  Jiang]{Wang2015}
{\sc \au{Wang, Q.}, \au{Yao, X.}, \au{Liu, H.}, \au{Qu{\'e}r{\'e}, D.} \&
  \au{Jiang, L.}} \yr{2015}  \at{Self-removal of condensed water on the legs of
  water striders}.  \jt{Proc. Natl. Acad. Sci.}  \bvol{112},  \pg{9247--9252}.

\bibitem[Zhu \& Gallaire(2016)]{Zhu2016}
{\sc \au{Zhu, L.} \& \au{Gallaire, F.}} \yr{2016}  \at{A pancake droplet
  translating in a hele-shaw cell: lubrication film and flow field}.  \jt{J.
  Fluid Mech.}  \bvol{798},  \pg{955--969}.

\end{thebibliography}

\end{document}